\documentclass[12pt]{article}
\usepackage[utf8x]{inputenc}

\usepackage{amsmath}
\usepackage{amsfonts}
\usepackage{amssymb}

\usepackage{todonotes}

\usepackage{graphicx, epstopdf}
\usepackage{graphicx,psfrag,epsf}

\usepackage{enumerate}

\usepackage[comma]{natbib}

\usepackage[ruled]{algorithm2e}

\usepackage{geometry}

\usepackage{microtype}

\usepackage{caption}
\usepackage{subcaption}

\usepackage{xcolor}

\newcommand{\E}{\operatorname{E}}
\newcommand{\Var}{\operatorname{Var}}

\newcommand{\argmax}{\operatornamewithlimits{argmax}}

\newcommand{\vect}[1]{\ensuremath{\boldsymbol{\mathbf{#1}}}}	
\newcommand{\matr}[1]{\ensuremath{\boldsymbol{\mathbf{#1}}}}  

\newtheorem{Def}{Definition}
\newtheorem{Theo}{Theorem}
\newtheorem{Res}{Result}
\newtheorem{Prop}{Proposition}
\newtheorem{Proo}{Proof}

\begin{document}

\def\spacingset#1{\renewcommand{\baselinestretch}%
{#1}\small\normalsize} \spacingset{1}

\title{Bayesian Spatial Inversion and \\Conjugate Selection Gaussian Prior Models}
\author{
       Henning Omre and Kjartan Rimstad \vspace{0.15cm} \\
\small Department of Mathematical Sciences\\
\small Norwegian University of Science and Technology\\
\small Trondheim, Norway \vspace{0.35cm}       }




\maketitle

\begin{abstract}

We introduce the concept of conjugate prior models for a given likelihood function in Bayesian spatial inversion. The conjugate class of prior models can be selection extended and still remain conjugate. We demonstrate the generality of selection Gaussian prior models, representing multi-modality, skewness and heavy-tailedness. For Gauss-linear likelihood functions, the posterior model is also selection Gaussian. The model parameters of the posterior pdf are explisite functions of the model parameters of the likelihood and prior models - and the actual observations, of course. Efficient algorithms for simulation of and prediction for the selection Gaussian posterior pdf are defined. Inference of the model parameters in the selection Gaussian prior pdf, based on one training image of the spatial variable, can be reliably made by a maximum likelihood criterion and numerical optimization. Lastly, a seismic inversion case study is presented, and improvements of $ 20$-$40\%$  in prediction mean-square-error, relative to traditional Gaussian inversion, are found.

\end{abstract}

\noindent%
\vfill

\newpage
\spacingset{1.45} 

\section{Introduction}
Inversion constitutes a challenge in many mathematical engineering problems. Observations from the variable of interest are often indirectly collected by some complex acquisition devise. The objective is naturally to predict the variable of interest based on the available observations. We consider spatial variables in this study and examples of inverse problems can be found in image analysis, remote sensing and geophysics. Inversion of seismic data is presented as a case study later in the paper.

The spatial variable of interest is $ \{ r(\vect{x}) ; \vect{x} \in \mathcal{D} \subset \mathcal{R}^m \} $ with $ r( \vect{x} ) \in \mathcal{R}  $ being the variable, having spatial reference $ \vect{x} $ running over the reference domain $ \mathcal{ D } \subset \mathcal{R}^m $ , which naturally has dimension $ m $ equal to one, two or three. The variable is discretized $ \{ r( \vect{x} ) ; \vect{x} \in \mathcal{L}_\mathcal{D} \} $ where $ \mathcal{L}_\mathcal{D} $ is a regular grid, of size $ n_r $, covering $ \mathcal{D} $ , and the spatial variable is represented by the $ n_r $-vector $ \vect{r} \in \mathcal{R}^{n_r} $. Assume further that a $ n_d $-vector of observations $ \vect{d} \in \mathcal{R}^{n_d}  $ related to the variable of interest is collected. The focus is on assessing $ \vect{r} $ given $ \vect{d} $, $ [ \vect{r} | \vect{d} ] $.

We phrase the assessment in a probabilistic setting, by using Bayesian inversion, see \citet{Tarantola2005},
\begin{align}
[ \vect{r} | \vect{d} ]  \rightarrow  f( \vect{r} | \vect{d} )  
               &= [ \int f( \vect{d} | \vect{r} ) f( \vect{r} ) d \vect{r} ]^{-1} \times f( \vect{d} | \vect{r} ) f( \vect{r} )\\
               &= \mbox{const} \times f( \vect{d} | \vect{r} ) f( \vect{r} )   \nonumber
\end{align}
where $ \vect{y} \rightarrow f( \vect{y} ) $ reads, the random variable $ \vect{y} $ is distributed according to the probability density function (pdf) $ f( \vect{y} ) $. The $ f( \vect{r} | \vect{d} ) $ is the posterior pdf being the ultimate solution of Bayesian inversion. The likelihood function $ f( \vect{d} | \vect{r} ) $ , being a function of $ \vect{r} $, represents the observation acquisition procedure, while the prior pdf $ f( \vect{r} ) $ summarizes prior information about the spatial variable of interest. The likelihood and prior models uniquely define the posterior model, although the integral in the normalizing constant usually is complicated to calculate.

Classical Bayesian inference, see \citet{casellaberger}, conserns estimation of model parameters, contrary to Bayesian inversion which is defined in a predictive setting. The classical approach focus on the posterior pdf of a low-dimensional vector of model parameters $ \vect{ \theta } $ given a set of observations $ \vect{y} $ , hence on $ f( \vect{\theta} | \vect{y} ) $. To assess this posterior pdf one must assume a likelihood function $ f( \vect{y} | \vect{ \theta} ) $ and specify a prior pdf $ f( \vect{ \theta} ) $. In order to avoid to calculate complex integrals, classical Bayesian inference has introduced the concept of conjugate classes of parametric prior pdfs. For a given likelihood function     $ f( \vect{y} | \vect{ \theta} ) $ , with a prior model $ f( \vect{ \theta} ) $  from the corresponding conjugate class of pdfs, the resulting posterior pdf $ f( \vect{ \theta} | \vect{y} ) $ will belong to the same conjugate class of pdfs. Consequently, the hyper-parameters of the posterior pdf will depend only on the hyper-parameters of the likelihood function and the prior pdf, and the actual observations, hence complex integral calculations are avoided.

In Bayesian spatial inversion the integral calculations are even more challenging than in classical Bayesian inference, since the variable of interest is of much higher dimensions than the model parameters. Fortunately, the elicitation of the prior model is simpler in Bayesian inversion than in Bayesian inference since the prior is on observable variables in the former while it is on model parameters in the latter. Hence Bayesian inversion may naturally be cast in an empirical Bayesian setting, see \citet{Efron1973} . 
In this study we consider Bayesian inversion and introduce the consept of conjugate classes of prior parametrised pdfs for a given class of likelihood functions. Further, we demonstrate that, for a given class of likelihood functions, the corresponding class of conjugate prior pdfs can be generalized by a selection mechanism. The resulting selection class of pdfs will also be conjugate with respect to the same class of likelihood functions. This construction makes it possible to define highly flexible classes of conjugate prior pdfs where the corresponding posterior pdfs can be determined by only the associated hyper-parameters and the actual observations. The assessment of the resulting posterior pdfs may require numerical or simulation schemes, but the closed form expressions for the posterior pdf makes it possible to design tailored efficient schemes. We demonstrate the selection conjugate consept on the familiar case of Gauss-linear likelihood functions and the conjugate class of spatial Gaussian pdfs.

Bayesian predictive inversion and Bayesian model inference can be combined to have hierarchical Bayesian inversion. It is, however, complicated to define general conjugate classes of prior models in this setting, see \citet{Roislien2006},
\citet{Arellano-Valle2009} and \citet{Branco2013}.
We present a brief discussion and evaluation of classical likelihood inference of the model parameters in the selection Gaussian prior pdf, from one available training image. Lastly, a case study of seismic inversion of real data is presented.

The developments of the selection Gaussian prior model is inspired by the early work on skewed pdfs in \citet{Azzalini1985} and \citet{Azzalini1996}, see also \citet{Genton2004} and \citet{Azzalini2013}. These models are generalized to spatial settings in  \citet{Kim2004}, \citet{Allard2007}, and \citet{Rimstad2012}. In the current study we use generalized selection sets as discussed in \citet{Arellano-Valle2004} and \citet{Arellano-Valle2006a}, to model spatial prior pdfs with marginal multi-modality. Also marginal skewness and heavy-tailedness may be represented. In spatial modelling, multi-modal spatial histograms occur if latent categorical variables are present. These variables may be lithologies in subsurface reservoir modelling or tissue classes in medical image analysis. Modelling multi-modal marginal characteristics with spatial continuity is challenging. If the observations are collected with spatial convolution and errors, as in subsurface acquisition and medical imaging, the inversion challenge is even larger.

The major contribution of the current paper is, however, the discussion of conjugate prior models in Bayesian spatial inversion and the demonstration that this conjugate characteristic is closed under activation of a selection mechanism. This result holds for all types of conjugate prior models. Moreover, we demonstrate the large potential of this result in modelling continuous spatial variables with multi-modal marginal distributions.

In the presentation $ f( \vect{y} ) $ denotes a pdf of the random variable $ \vect{y} $, while $ F( \vect{y} \in \mathcal{B} ) $ denotes the probability that the random variable is in the sub-set $ \mathcal{B} $ of its sample space. For the Gaussian random $ n $-vector    $ \vect{y} $ we write,
\begin{align}
\vect{y} \rightarrow f( \vect{y}) &= \phi_n ( \vect{y} ; \vect{ \mu }, \matr{ \Sigma} ) \\
                                              &= [2\pi]^{-n/2} |\matr{\Sigma}|^{-1/2} 
                                                                          \exp\{ - \frac{1}{2} [\vect{y}-\vect{\mu}]^T \matr{\Sigma}^{-1} [\vect{y}-\vect{\mu}]\}    \nonumber     \\   
F( \vect{y} \in \mathcal{B} ) &= \Phi_n ( \mathcal{B} ; \vect{\mu}, \matr{\Sigma)} = \int_{\mathcal{B}} \phi_n( \vect{u} ; \vect{\mu}, \matr{\Sigma}) d\vect{u} \nonumber
\end{align}
Note that from a computational point of view $ |\matr{\Sigma}|^{-1} $ and $ \matr{\Sigma}^{-1} $ are demanding, and so is simulation from and  calculation of the sub-set $ \mathcal{B} $ probability of an arbitrary high-dimensional Gaussian pdf. The former two are widely studied, however, while the latter has drawn much less attention except for \citet{Genz2009}. Later, we suggest efficient algorithms for simulation from and calculating of these sub-set $ \mathcal{B} $ probabilities.
We also use the notation $ \vect{i}_n $ for a unit $ n$-vector and $ \matr{I}_n $ for a identity $ (n \times n)$-matrix, while $ I(A) $ is an indicator function taking value $1$ if $ A$ is true and $ 0 $ otherwise.

In Section 2, Bayesian spatial inversion is discussed, and a selection extended conjugate class of prior pdfs for a given class of likelihood functions is defined. The conjugate class of selection Gaussian prior pdfs is developed and discussed in detail. Expressions for the model parameters of the corresponding posterior pdf are developed, and simulation algorithms for assessing the posterior pdf are presented. The flexibility of the class of prior pdfs is illustrated by several examples.
Section 3 contains a discussion of model parameter inference and the development of maximum likelihood estimators for the model parameters of the conjugate class of selection Gaussian prior pdfs, based on one training realization of the spatial variable. A small empirical study demonstrates the consistency of the estimator as the size of the training variable increase. In Section 4, a case study based on real seismic data along a well profile from the Alvheim field in the North Sea is presented, see also \citet{karimi:R1}. Comparisons with regular Bayesian Gaussian inversion are made.  Lastly, in Section 5, the conclusions of the study are forwarded.

\section{Bayesian Spatial Inversion}

The focus is on prediction of a spatial variable discretized into the $ n_r $-vector $ \vect{r} $, based on the observations represented in the $ n_d $-vector $ \vect{d} $. We phrase the prediction as Bayesian inversion, see Expression 1, which requires that the likelihood function $ f( \vect{d} | \vect{r} ) $ is given, and that the prior pdf $ f( \vect{r} ) $ is specified.  Hereby, the corresponding posterior pdf $ f( \vect{r} | \vect{d} ) $ is defined. Inspired by the consept of conjugate prior pdfs in traditional Bayesian inference we present the following definition,
\begin{Def}[Conjugate class of prior pdfs] 
Consider Bayesian inversion,
\begin{align*}
f( \vect{r} | \vect{d} ; \vect{\theta}_{r|d})  
               = \mbox{const} \times f( \vect{d} | \vect{r} ; \vect{\psi}_d ) f( \vect{r} ; \vect{\theta}_r )
\end{align*}
with likelihood function $ f( \vect{d} | \vect{r} ; \vect{\psi}_d ) $ in a parametrized pdf class $ \mathcal{L}_{\psi} $ and
prior pdf $ f( \vect{r} ; \vect{\theta}_r ) $ in a parametrized pdf class $ \mathcal{P}_{\theta} $. If the associated posterior pdf 
$ f( \vect{r} | \vect{d} ; \vect{\theta}_{r|d}) $ also is in the pdf class $ \mathcal{P}_{\theta} $, then the pdf class 
$ \mathcal{P}_{\theta} $ is termed a conjugate class with respect to the likelihood function class $ \mathcal{L}_{\psi} $. The 
posterior model parameters $ \vect{\theta}_{r|d} $ will be a function of $ [ \vect{\psi}_d , \vect{\theta}_r , \vect{d} ] $.
\label{Def:conj}
\end{Def}
For continuous spatial variables, the class of Gaussian prior pdfs is known to be a conjugate class with respect to Gauss-linear likelihood functions. Hence, if the observations are collected through a linear forward model with additive Gaussian errors, and the prior pdf is specified to be Gaussian, then the posterior pdf will also be Gaussian. This characteristic is the basis for kriging prediction and conditional simulation in geostatistics, see \citet{Chiles2012}. Moreover, for event spatial variables the Poisson prior pdf is conjugate with respect to thinning likelihood functions, while for mosaic spatial variables the Markov prior pdf is conjugate with respect to conditionally independent single-site response likelihood functions. These conjugate characteristics are of course the major reason for the frequent use of these spatial models.
In the next section we define an extended class of prior pdfs based on a selection consept, and demonstrate that this consept can be used to construct an extended conjugate class of prior pdfs. 

\subsection{Generalization by Selection}

Consider the previously defined spatial variable represented by the $ n_r $-vector $ \vect{r} $ with prior pdf $ f( \vect{r} ) $. Extend the dimension by an auxiliary random $ n_\nu $-vector  $ \vect{\nu} \in \mathcal{R}^{n_\nu} $, such that,
\begin{align}
\left[
\begin{array}{c}
\vect{r}  \\
\vect{\nu}
\end{array}  \right]
\rightarrow
f \left(  \left[
\begin{array}{c}
\vect{r}  \\
\vect{\nu}
\end{array}  \right]  \right)
=
f ( \vect{\nu} | \vect{r} ) 
f ( \vect{r} )
\end{align}
with arbitrary chosen pdf $ f( \vect{\nu} | \vect{r} ) $, and denote the pdf $ f( \vect{r} ) $ the basis-pdf. Consider an arbitrary sub-set $ \mathcal{A} \subset \mathcal{R}^{n_\nu} $ and define the associated random selection $ n_r $-vector $ \vect{r}_A $ by,
\begin{align}
\vect{r}_A = [ \vect{r} | \vect{\nu} \in \mathcal{A} ]  \rightarrow f( \vect{r}_A ) &= f( \vect{r} | \vect{\nu} \in \mathcal{A} ) \\ 
                &= [ F( \vect{\nu} \in \mathcal{A} ) ]^{-1} \times  F( \vect{\nu} \in \mathcal{A} | \vect{r} ) f( \vect{r} ) \nonumber
\end{align}
Note in particular, that $ f( \vect{r}_A )  =  f( \vect{r} ) $ if we define $ f( \vect{\nu} | \vect{r}) = f( \vect{\nu} ) $, which of course is the extreme choice of independence between $ \vect{r} $ and $ \vect{\nu} $. Based on this selection consept we define,
\begin{Def}[Selection extension of prior pdfs] Consider a prior basis-pdf $ f ( \vect{r} ; \vect{\theta}_r ) $ in a \linebreak parametrized pdf class
$ \mathcal{P}_{\theta} $ and define auxiliary variable $ \vect{\nu} \in \mathcal{R}^{n_\nu} $ related to $ \vect{r} $ by pdf 
$ f ( \vect{\nu} | \vect{r} , \vect{\kappa}_{\nu} ) $ in a parametrized pdf class $ \mathcal{E}_{\kappa} $. Specify further a selection set $ \mathcal{A} \subset \mathcal{R}^{n_\nu} $.  Define the selection variable,
\begin{align*}
\vect{r}_A = [ \vect{r} | \vect{\nu} \in \mathcal{A} ]  \rightarrow f( \vect{r}_A ) 
&= f( \vect{r} | \vect{\nu} \in \mathcal{A} ; \vect{\theta}_r , \vect{\kappa}_\nu ) \\ 
     &= [ F( \vect{\nu} \in \mathcal{A}; \vect{\theta}_r , \vect{\kappa}_\nu ) ]^{-1} \times  
     F( \vect{\nu} \in \mathcal{A} | \vect{r}; \vect{\kappa}_\nu ) f( \vect{r} ;\vect{\theta}_r ) 
\end{align*}
with pdf $ f( \vect{r} | \vect{\nu} \in \mathcal{A} ; \vect{\theta}_r , \vect{\kappa}_\nu ) $ in the parametrized selection extended pdf class $ \mathcal{S}_\mathcal{A} [ \mathcal{P}_\theta  \times \mathcal{E}_\kappa ] $.
\label{Def:sel_ext}
\end{Def}
The selection extension can be made for any basis-pdf class for arbitrary auxiliary variables with associated selection sets. The class of selection Gaussian pdfs with $ f ( \vect{r} ) $ from the Gaussian class and $ f ( \vect{\nu} | \vect{r} ) $ being Gauss-linear with associated selection sets, hence $ [ \vect{r} , \vect{\nu} ] $ being  jointly Gaussian, is thoroughly discussed in \citet{Arellano-Valle2006a}. We define this class of selection Gaussian pdf by a Gaussian basis-pdf
\begin{align*}
\vect{r} \rightarrow f( \vect{r} ) = \phi_{ n_r} (\vect{r} ; \vect{\mu}_r , \matr{\Sigma}_r ) 
\end{align*}
with the expectation $ n_r $-vector $ \vect{\mu}_r $ and the covariance $ ( n_r \times n_r ) $-matrix $ \matr{\Sigma}_r $. 
The auxiliary $ n_\nu $-vector $ \vect{\nu} $ is defined as,
\begin{align*}
[ \vect{\nu} | \vect{r} ] \rightarrow f( \vect{\nu} | \vect{r} ) 
= \phi_{n_\nu} ( \vect{\nu}; \vect{\mu}_{\nu|r} , \matr{\Sigma}_{ \nu|r} )
\end{align*}
with the conditional expectation $ n_\nu $-vector being linear in $ \vect{r} $, 
$ \vect{\mu}_{\nu|r} = \vect{\mu}_\nu + \matr{\Gamma}_{\nu|r} ( \vect{r} - \vect{\mu}_r ) $ with expectation $ n_\nu $-vector $ \vect{\mu}_\nu $,
coupling $ ( n_\nu \times n_r ) $-matrix $ \matr{\Gamma}_{\nu|r} $,
 and the conditional covariance $ ( n_r \times n_\nu ) $-matrix $ \matr{\Sigma}_{\nu|r} $. 
Hence $ [ \vect{\nu} | \vect{r} ] $ is Gauss-linear, and since 
$ \vect{r} $ is Gaussian, the joint $ (n_r + n_\nu) $-vector $ [ \vect{r} , \vect{\nu} ] $ is Gaussian. 
By enforcing the selection $ \vect{ \nu} \in \mathcal{A} \subset \mathcal{R}^{n_\nu} $, 
we obtain the selection Gaussian $ n_r $-vector $ \vect{r}_A $,
\begin{align}  \label{eq:selGaus}
\vect{r}_A = [ \vect{r} | \vect{\nu} \in \mathcal{A} ] \rightarrow 
f ( \vect{r}_A ) &= f ( \vect{r} | \vect{\nu} \in \mathcal{A} ) \\
&= [ \Phi_{n_\nu } ( \mathcal{A} ; \vect{\mu}_\nu , \matr{\Sigma}_\nu ) ]^{-1} \times
\Phi_{n_\nu} ( \mathcal{A} ; \vect{\mu}_{\nu|r} , \matr{\Sigma}_{\nu|r} )  
\phi_{n_r} ( \vect{r} ; \vect{\mu}_r , \matr{\Sigma}_r ) \nonumber
\end{align}
where the marginal covariance $ ( n_\nu \times n_\nu ) $-matrix is 
$ \matr{\Sigma}_\nu = \matr{\Gamma}_{\nu|r} \matr{\Sigma}_r \matr{\Gamma}_{\nu|r}^T + \matr{\Sigma}_{\nu|r} $.
All valid sets of model parameters 
$ ( \vect{\mu}_r , \matr{\Sigma}_r , \vect{\mu}_\nu , \matr{\Gamma}_{\nu|r} , \matr{\Sigma}_{\nu|r}, \mathcal{A} ) $, 
define the class of selection Gaussian pdfs. Note, that by assigning $ \matr{\Gamma}_{\nu|r} $ a null-matrix or setting $ \mathcal{A} = \mathcal{R}^{n_\nu} $, the selection Gaussian class of pdfs is identical to the Gaussian one.

The likelihood function of the actual observations $ \vect{d}^o $,   $ f( \vect{d}^o | \vect{r} ) $,  is a function of $ \vect{r} $ and only dependent on the observation acquisition procedure, independent on choice of prior pdf. Hence, the posterior pdf based on a prior selection pdf, can be expressed as,
\begin{align}
     [ \vect{r}_A | \vect{d}^o ]  \rightarrow f( \vect{r}_A | \vect{d}^o )  &= \mbox{const} 
                                                \times f( \vect{d}^o | \vect{r}_A ) f( \vect{r}_A)  \\
                   &= \mbox{const}_1 \times f( \vect{d}^o | \vect{r}) f( \vect{r} | \vect{\nu} \in \mathcal{A} ) \nonumber \\
              &= \mbox{const}_2 \times f( \vect{d}^o | \vect{r} ) F( \vect{\nu} \in \mathcal{A} | \vect{r} ) f( \vect{r} ) \nonumber  \\
                &= \mbox{const}_3 \times F(  \vect{\nu} \in \mathcal{A} | \vect{r} , \vect{d}^o ) f( \vect{r} | \vect{d}^o ) \nonumber \\
                  &= [ F( \vect{\nu} \in \mathcal{A} | \vect{d}^o) ]^{-1} 
\times F( \vect{\nu} \in \mathcal{A} | \vect{r}, \vect{d}^o ) 
                                                f ( \vect{r} | \vect{d}^o ) \nonumber
\end{align}
which relies on the conditional independence relation $ f( \vect{\nu}, \vect{d} | \vect{r} ) = f( \vect{\nu} | \vect{r} ) f( \vect{d} | \vect{r} ) $. Note that the posterior pdf corresponds to the selection pdf with basis-pdf being the conditional pdf $ f( \vect{r} | \vect{d} ) $, which provides the following theorem,
\begin{Theo}[Selection extended conjugate class of prior pdfs] Consider a likelihood function in parametrized pdf class $ \mathcal{L}_\psi $ and a prior pdf in the parametrized pdf class $ \mathcal{P}_\theta $ .  According to Definition \ref{Def:conj} - let the pdf class 
$ \mathcal{P}_\theta $ be a conjugate class with respect to likelihood function class $ \mathcal{L}_\psi $. According to Definition \ref{Def:sel_ext} - define the associated selection extended pdf class $ \mathcal{S}_\mathcal{A} [ \mathcal{P}_\theta  \times \mathcal{E}_\kappa ] $ based on auxiliary pdf class $ \mathcal{E}_\kappa $ and selection set $ \mathcal{A} $.  Then the pdf class 
$ \mathcal{S}_\mathcal{A} [ \mathcal{P}_\theta  \times \mathcal{E}_\kappa ] $ is a conjugate class with respect to the likelihood function class $ \mathcal{L}_\psi $ for all pdf classes $ \mathcal{E}_\kappa $ and selection sets $ \mathcal{A} $. The conjugate characteristics of a prior pdf class $ \mathcal{P}_\theta $ is closed under selection extension.
\label{Theo:sel_ext_conj}
\end{Theo}
This closedness property for conjugate pdf classes is very general and applies to continuous, event and mosaic spatial variables. Moreover, it may be used in traditional Bayesian inference.
For continuous spatial variables with a Gauss-linear likelihood function, prior pdfs from the class of Gaussian pdfs is known to be conjugate. According to the results above, also the selection Gaussian pdf for any arbitray  auxiliary extension and  selection set, will define a conjugate class of pdfs with repect to a Gauss-linear likelihood function. In the following  sub-sections we will explore this opportunity to define more flexible prior pdfs in Bayesian spatial inversion with observations collected through a Gauss-likelihood function.

\subsection{Likelihood Model}
We limit the likelihood function to be from the Gauss-linear class,
\begin{align}
[ \vect{d} | \vect{r} ] = \matr{H} \vect{r} + \vect{e}_{d|r}  \rightarrow  f( \vect{d} | \vect{r} ) = \phi_{n_d} ( \vect{d} ; \matr{H} \vect{r}, \matr{\Sigma}_{d|r} )
\end{align}
where $ \matr{H} $ is an observation acquisition $ ( n_d \times n_r ) $-matrix and $ \vect{e}_{d|r} $ is a centred Gaussian error $ n_d $-vector with covariance  $ ( n_d \times n_d ) $-matrix $ \matr{\Sigma}_{d|r} $, independent of $ \vect{r} $. Hence the model parameters are $ \vect{\theta}_l = ( \matr{H}, \matr{\Sigma}_{d|r} ) $. There are no constraint on the matrix $ \matr{H} $. It may be binary,  selecting only a sub-set of elements in the vector $ \vect{r} $ or it could represent convolution by averaging over elements in the vector $ \vect{r} $. Also numerical differentiation or integration can be captured by $ \matr{H} $.  In the case study we use a convolved, contrast-linearized approximation to the wave equation to model acquisition of seismic data, see \citet{buland:185}.

\subsection{Prior Model}\label{Prior_Mod}
The spatial variable of interest is represented in the vector $ \vect{r} $. We define the prior basis-pdf $ f( \vect{r} ) $ to be a spatially stationary Gaussian pdf,
\begin{align}
\vect{r} \rightarrow  f( \vect{r} ) = \phi_{n_r} ( \vect{r} ; \mu \vect{i}_{n_r}, \sigma^2 \matr{C})
\end{align}
where the scalars $ ( \mu, \sigma^2) $ are the stationary expectation and variance, respectively, while the spatial correlation $ ( n_r \times n_r) $-matrix $ \matr{C} $ is defined by the spatial translation invariant correlation function $ \rho( \vect{\tau}); \vect{\tau} \in \mathcal{R}^m $ .  This pdf is spatially stationary in the sense that the marginal pdfs $ f(r_i) = \phi_1 ( r_i ; \mu, \sigma^2) ; i=1, \dots ,n_r $ are all identical. Moreover, it exhibits ergodisity since $ f( r_i , r_j ) \rightarrow f(r_i ) f( r_j ) $  as $ | \vect{x}_i - \vect{x}_j | \rightarrow \infty $, which entails that consistent estimates of the model parameters can be obtained.

We use this Gaussian pdf as basis-pdf to define a selection pdf, with the auxiliary  $ n_r $-vector $ \vect{\nu} $ extension,
\begin{align}
 [ \vect{\nu} | \vect{r} ]  =  \gamma \sigma^{-1} [ \vect{r} - \mu \vect{i}_{n_r}] + \vect{e}_{\nu | r}  
                                 \rightarrow  f( \vect{\nu} | \vect{r} ) 
       &= \phi_{n_r} ( \vect{\nu} ; \gamma \sigma^{-1} [ \vect{r} - \mu \vect{i}_{n_r} ], [1- \gamma^2] \matr{I}_{n_r} ) 
 \nonumber \\
&= \prod_{i=1}^{n_r} \phi_1 ( \nu_i ; \gamma \sigma^{-1} [ r_i - \mu ] , [ 1 - \gamma^2 ] )  \nonumber
\end{align}
where $ \gamma \in [-1,1] \subset \mathcal{R} $ is a coupling parameter while $ \vect{e}_{ \nu | r} $ is a centred Gaussian $ n_r $-vector with independent elements with variance $ [1 - \gamma^2 ] $, independent of $ \vect{r} $. The extended variable becomes jointly Gaussian,
\begin{align}
\left[
\begin{array}{c}
\vect{r} \\  \vect{\nu}
\end{array} \right]
\rightarrow  
f \left( 
\left[
\begin{array}{c}
\vect{r} \\  \vect{\nu}
\end{array} \right] \right)
= \phi_{2n_r} \left( 
\left[
\begin{array}{c}
\vect{r} \\  \vect{\nu}
\end{array} \right]
; \left[ \begin{array}{c} 
\mu \vect{i}_{n_r} \\  0 \vect{i}_{n_r}
\end{array}  \right] , 
\left[
\begin{array}{cc}
\sigma^2 \matr{C}  & \gamma \sigma \matr{C} \\
\gamma \sigma \matr{C} & \gamma^2 \matr{C} + [1 - \gamma^2] \matr{I}_{n_r}
\end{array}  \right]
 \right)
\end{align}
and consists of two grid-discretized spatial variables with variances $ \sigma^2 $ and $ 1 $ respectively, with inter-correlation $ \gamma \sigma $ and with identical spatial correlation function $ \rho ( \vect{\tau} ) $.

Define the selection set $  \mathcal{A} = \cup_{i=1}^{n_r} \mathcal{A}_i \subset \mathcal{R}^{n_r} $ with $ \mathcal{A}_i = \mathcal{A}_j ; i,j = 1, \dots ,n_r $ , hence identical selection sets for each component in $ \vect{r} $. The corresponding spatial selection Gaussian pdf, which belong to the class of selection Gaussian pdfs, see Expression \ref{eq:selGaus}, is defined as,
\begin{align}
\vect{r}_A  =  [ \vect{r} | \vect{\nu} \in \mathcal{A} ]  \rightarrow  &f( \vect{r}_A ) = f( \vect{r} | \vect{\nu} \in \mathcal{A} ) \\  
                  &=[ F( \vect{\nu} \in \mathcal{A} ) ]^{-1} \times F ( \vect{\nu} \in \mathcal{A} | \vect{r} ) f ( \vect{r} ) \nonumber  \\
                 &= [ \Phi_{n_r} ( \cup_{i=1}^{n_r} \mathcal{A}_i ; 0 \vect{i}_{n_r} , \gamma^2 \matr{C} + [1-\gamma^2]               
                 \matr{I}_{n_r})]^{-1} \nonumber \\
                    &\times \prod_{i=1}^{n_r} \Phi_1 ( \mathcal{A}_i ; \gamma \sigma^{-1} [r_i - \mu], [1-\gamma^2]) 
                  \mbox{  }   \phi_{n_r} ( \vect{r} ; \mu \vect{i}_{n_r} , \sigma^2 \matr{C} ) \nonumber
\end{align}
with the actual model parameters  $ \vect{\theta}_p = ( \mu, \sigma^2, \gamma, \rho( \vect{\tau}), \mathcal{A}_i) $.

Note in particular that the prior model for the spatial variable of interest, represented by the selection Gaussian pdf on the vector $ \vect{r} $, is subject to the grid $ \mathcal{L}_{\mathcal{D}} $. The selection Gaussian consept breaks down when the grid size tends to zero by infilling of the grid, see \citet{Minozzo2012} and \citet{Rimstad2012}.  This lack of generality limits the use of the model, of course, but in many applications like image analysis, remote sensing and geophysics, there is a natural choice of grid $ \mathcal{L}_{\mathcal{D}} $ due to the observation  acquisition procedure. Moreover, this limitation is sheared by the categorical Markov random field model, see \citet{besag1974} and \citet{Kaiser2002}, which has proven immensely useful in many applications.

The selection Gaussian class of pdfs can be shown to be closed under marginalization,
 see \citet{Arellano-Valle2006a}, and the uni-variate marginal pdf is,
\begin{align}
  r_{Ai} = [ r_i | \vect{\nu} \in \mathcal{A} ]  \rightarrow  &f( r_{Ai} ) = f( r_i | \vect{\nu} \in \mathcal{A} )  \\
                  &= [ F ( \vect{\nu} \in \mathcal{A} ) ]^{-1}  \times F ( \vect{\nu} \in \mathcal{A} | r_i ) f( r_i ) \nonumber  \\
                   &=[ \Phi_{n_r} ( \cup_{i=1}^{n_r} \mathcal{A}_i ; 0 \vect{i}_{n_r} , \gamma^2 \matr{C} + [1-\gamma^2]               
                 \matr{I}_{n_r})]^{-1} \nonumber \\
                 &\times \Phi_{n_r} ( \cup_{i=1}^{n_r} \mathcal{A}_i ; \gamma \sigma^{-1} \vect{c}_i (r_i - \mu ),
                 \gamma^2 [ \matr{C} - \vect{c}_i \vect{c}_i^T ] + [1 - \gamma^2] \matr{I}_{n_r} ) \nonumber \\
                 &\times \phi_1 ( r_i; \mu , \sigma^2 ) \nonumber 
\end{align}
which is a selection Gaussian pdf with $ \vect{c}_i $ being the $ i$'th column $ n_r $-vector of the correlation matrix $ \matr{C} $.   Note, however, that the two first moments of the marginal pdf, $ \E(r_{Ai}) $  and $ \Var(r_{Ai}) $ do not have nice closed form expressions. All lower dimensional marginal pdfs will also be selection Gaussian pdfs.

All marginal pdfs $ f( r_{Ai}) ; i=1, \dots , n_r $ are dependent on all $ n_r $ elements of the auxiliary variable $ \vect{\nu} $, not only $ \nu_i $ . This dependence causes  the selection Gaussian pdf to be defined subjective to the grid $ \mathcal{L}_{\mathcal{D}} $.  This coupling is contrary to Gaussian marginal pdfs where all dependence of other dimensions are integrated out.  The dependence will decline with distance, however, since the spatial correlation function $ \rho( \vect{\tau}) $ will tend towards zero with increasing $ | \vect{\tau} | $. Consequently all marginal pdfs will be equal, due to symmetry of the regular grid $ \mathcal{L}_\mathcal{D} $, except for edge effects of the grid. The decline of the spatial correlation with distance will also make the bi-variate $ f( r_{Ai} , r_{Aj} ) $ tend towards $  f( r_{Ai} )  f( r_{Aj} ) $, hence independence,  as the distance $ | \vect{\tau}_{ij} | = | \vect{x}_i - \vect{x}_j | $ increase. Consequently, the selection Gaussian pdf exhibit approximate stationarity and ergodicity in the sense defined above.  These characteristics make the selection Gaussian pdf suitable as prior pdf in Bayesian spatial inversion.

The prior selection Gaussian pdf will naturally be inspected by simulation. Simulation is performed sequentially by first generating a realization of the auxiliary variable $ \vect{\nu}^s \in \mathcal{A} $ and thereafter generating the realization $ \vect{r}_A^s = [ \vect{r} | \vect{\nu}^s ]^s $.  Since the joint variable $ [ \vect{r}, \vect{\nu} ] $ is Gaussian, many efficient algorithms are available, see Supplement \ref{sec:app_sample}.

\begin{figure}
 \centering
\includegraphics[height=0.7\textheight]{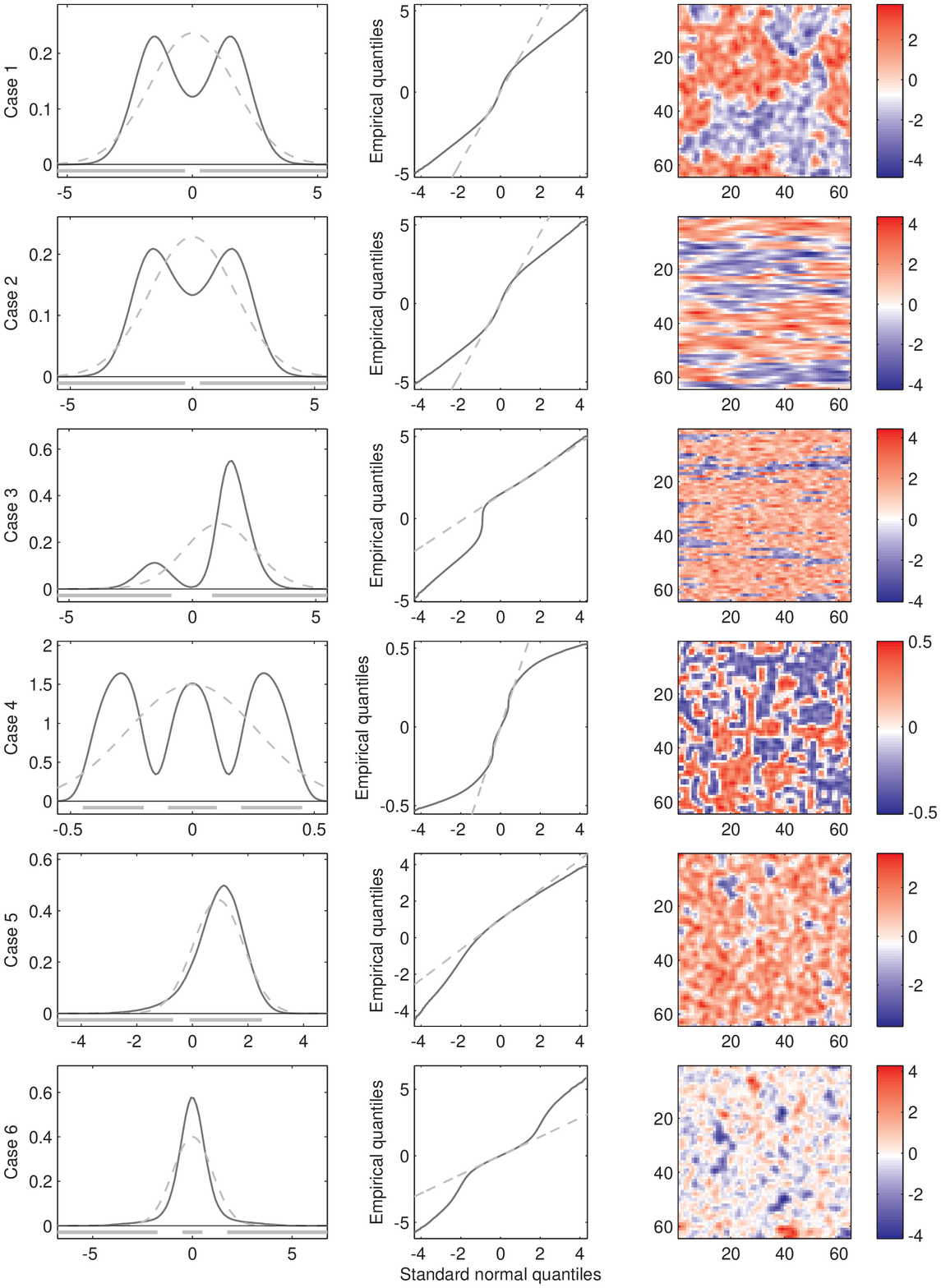}
 \caption{First column: marginal distribution of selection Gaussian random field ( solid black), standard normal distribution ( dashed gray), and selection sets on auxiliary random field on axis ( solid gray). Second column: quantile-quantile plot of marginal selection Gaussian random field versus theoretical quantiles from the Gaussian distribution. Third column: realization from selection Gaussian random field.}
 \label{fig:synthetic_marginals}
\end{figure}

The variety of the prior selection Gaussian model is exhibited in Figure \ref{fig:synthetic_marginals} which is based on the model parameters listed in Table \ref{tbl:synthetic_param} with an anisotropic second-order exponential correlation function with anisotropy factor $ ( d_h , d_v ) $. A more detailed discussion of the example is given in Supplement \ref{sec:app_exPrior}. The spatial variable is represented on a $ (64 \times 64) $-grid. The anisotropy factors vary, and so do the correlation and the selection sets for the auxiliary variable. We observe a large variety of prior spatial models, all of them approximately stationary and ergodic in the sense discussed above. The prior models have marginal distributions that can be multi-modal, skewed or heavy-tailed, or a combination of these features. The computer demand for generating one such realization is typically a couple of minutes on a regular laptop computer.

\vspace{0.5cm}
\begin{table}
\caption{Model parameters for six cases, with $\mu = 0 $ and  $\sigma^2=1$ for all cases.}
\begin{center}
\begin{small}
\begin{tabular}{c|ccccl }
Case & $\gamma$ & $d_h$ & $d_v$ & $A_i$ & description \\ \hline
1    & 0.8000   & 2.0   & 2.0  & $(\infty, -0.3] \cup [0.3, \infty)$ & sym. bimodal iso. \\ 
2    & 0.6500   & 6.0   & 0.85  & $(\infty, -0.3] \cup [0.3, \infty)$ & sym. bimodal aniso. \\
3    & 0.9250   & 2.0   & 0.60  & $(\infty, -0.85] \cup [0.8, \infty)$& asym. bimodal aniso. \\
4    & 0.9995   & 3.0   & 3.0  & $[-0.45, -0.2] \cup [-0.1, 0.1] \cup [0.2, 0.45]$ & sym. trimodal iso.\\
5    & 0.7000   & 2.0   & 2.0  & $(\infty, -0.7] \cup [-0.1, 2.5]$    & asym. unimodal iso. \\
6    & 0.7000   & 2.0   & 2.0  & $(\infty, -1.75] \cup [-0.5, 0.5] \cup [1.75, \infty)$ & sym. heavy tailed iso. \\
\end{tabular}
\end{small}
\end{center}
\label{tbl:synthetic_param}
\end{table}

\subsection{Posterior Model} \label{Post_Mod}

The posterior pdf is uniquely defined by the likelihood function and the prior pdf. With a likelihood function from the Gauss-linear class and a prior pdf from the selection Gaussian class, the posterior pdf will also, due to Theorem \ref{Theo:sel_ext_conj}, be from the selection Gaussian class. The selection Gaussian class of prior pdfs is conjugate with respect to Gauss-linear likelihood functions. Hence the model parameters of the posterior pdf are analytically tractable based on the model parameters of the likelihood and prior models and the actual observations. The joint pdf is,
\begin{align}
&\left[
\begin{array}{c}
\vect{r}  \\
\vect{\nu} \\
\vect{d}
\end{array} \right]
\rightarrow
f \left( \left[
\begin{array}{c}
\vect{r}  \\
\vect{\nu} \\
\vect{d}
\end{array} \right]  \right) \\
&=
\phi_{2n_r + n_d} \left(  \left[
\begin{array}{c}
\vect{r}  \\
\vect{\nu} \\
\vect{d}
\end{array}  \right]
;  \left[
\begin{array}{c}
\mu \vect{i}_{n_r}  \\
0 \vect{i}_{n_r} \\
\mu \matr{H} \vect{i}_{n_r}
\end{array} \right]
, \left[
\begin{array}{ccc}
\sigma^2 \matr{C} & \gamma \sigma \matr{C} & \sigma^2 \matr{C} \matr{H}^T \\
\gamma \sigma \matr{C}  &  \gamma^2 \matr{C} + [1-\gamma^2] \matr{I}_{n_r} & \gamma \sigma \matr{C} \matr{H}^T \\
\sigma^2 \matr{H} \matr{C} & \gamma \sigma \matr{H} \matr{C}  &  \sigma^2 \matr{H} \matr{C} \matr{H}^T + \matr{\Sigma}_{d|r}
\end{array}  \right]  \right)  \nonumber
\end{align}
and one may demonstrate that $ [\vect{\nu} , \vect{d} | \vect{r} ] $ are conditionally independent. Note also that the joint $ [ \vect{r} , \vect{d} | \vect{\nu} \in \mathcal{A} ] $ will be selection Gaussian, and so will the two marginals $ [ \vect{r} | \vect{\nu} \in \mathcal{A} ] $ and $ [ \vect{d} | \vect{\nu} \in \mathcal{A} ] $. Hence the marginal pdf of the observations will be dependent on the actual prior model, which the likelihood model will not. The focus of the study is on the posterior 
$ [ \vect{r} | \vect{d} ,\vect{\nu} \in \mathcal{A} ] $ which will be selection Gaussian as well, see Appendix \ref{app:selGauss}.

From Theorem  \ref{Theo:sel_ext_conj} and standard Gaussian theory one has,
\begin{align}
[\vect{r}_A | \vect{d} ] =  [ \vect{r} | \vect{\nu} \in \mathcal{A}, \vect{d} ]  \rightarrow  &f( \vect{r}_A | \vect{d}) 
              = f( \vect{r} | \vect{\nu} \in \mathcal{A} , \vect{d}) \\ 
              &= [ F( \vect{\nu} \in \mathcal{A}| \vect{d} ) ]^{-1} \times  
                    F ( \vect{\nu} \in \mathcal{A} | \vect{r}, \vect{d} ) f ( \vect{r}| \vect{d} ) \nonumber \\
                 &= [ \Phi_{n_r} ( \cup_{i=1}^{n_r} \mathcal{A}_i ;  \vect{\mu}_{\nu|d} , \matr{\Sigma}_{\nu | d} )]^{-1} \nonumber \\
                    &\times  \Phi_{n_r} (\cup_{i=1}^{n_r} \mathcal{A}_i ; \vect{\mu}_{\nu|rd}, \matr{\Sigma}_{\nu |rd} ) \nonumber \\
                    &\times \phi_{n_r} ( \vect{r} ; \vect{\mu}_{r|d}  , \matr{\Sigma}_{r|d} ) \nonumber
\end{align}
with 
\begin{align*}
\left[
\begin{array}{c}
\vect{\mu}_{r|d} \\  \vect{\mu}_{\nu|d}
\end{array}  \right]
= \left[
\begin{array}{c}
 \mu \vect{i}_{n_r} \\ 0 \vect{i}_{n_r}
\end{array}  \right]
+ \left[
\begin{array}{c}
\sigma^2 \matr{C} \matr{H}^T  \\  \gamma \sigma \matr{C}  \matr{H}^T 
\end{array}  \right]
\left[
\sigma^2 \matr{H} \matr{C} \matr{H}^T + \matr{\Sigma}_{d|r}
\right]^{-1}
\left[
\vect{d} - \mu \matr{H}  \vect{i}_{n_r}
\right] 
\end{align*}
\begin{align*}
\left[
\begin{array}{cc}
\matr{\Sigma}_{r|d}  &  \matr{\Gamma}_{r \nu|d} \\
\matr{\Gamma}_{\nu r|d} & \matr{\Sigma}_{\nu | d}
\end{array} \right]
=  &\left[
\begin{array}{cc}
\sigma^2 \matr{C}  &  \gamma  \sigma \matr{C}  \\
\gamma \sigma \matr{C}  &  \gamma^2 \matr{C} + [1 - \gamma^2 ] \matr{I}_{n_r}
\end{array}  \right] \\
&- \left[
\begin{array}{c}
\sigma^2 \matr{C} \matr{H}^T \\
\gamma \sigma \matr{C} \matr{H}^T 
\end{array} \right]
\left[
\sigma^2 \matr{H} \matr{C} \matr{H}^T + \matr{\Sigma}_{d|r}
\right]^{-1}
\left[
\begin{array}{cc}
\sigma^2 \matr{H} \matr{C}  &  \gamma \sigma \matr{H}  \matr{C}
\end{array}  \right]
\end{align*}
\begin{align*}
\vect{\mu}_{\nu | rd} &= \vect{\mu}_{\nu | d} + \matr{\Gamma}_{vr|d} \matr{\Sigma}_{r|d}^{-1} [\vect{r} - \vect{\mu}_{r|d}] \\
\matr{\Sigma}_{\nu | rd} &= \matr{\Sigma}_{\nu |d} - \matr{\Gamma}_{\nu r|d} \matr{\Sigma}_{r|d}^{-1} \matr{\Gamma}_{r \nu | d}
\end{align*}
  This posterior pdf will of course be spatially non-stationary due to conditioning on the observations $ \vect{d} $. The pdf will, however, be in the class of selection Gaussian pdfs, see Expression \ref{eq:selGaus}, and hence be closed under marginalization and conditioning, with the corresponding model parameters analytically tractable. Assessment of the posterior pdf is usually made by simulation of realizations and locationwise prediction with associated precision intervals.

Simulation of realizations from the posterior pdf is made sequentially by first to generate a realization $ [ \vect{\nu}^s \in \mathcal{A} | \vect{d} ] $ and thereafter to generate a realization $ [ \vect{r}_A | \vect{d} ]^s = [ \vect{r} | \vect{\nu}^s, \vect{d} ]^s $. Since the joint variable $ [ \vect{r}, \vect{\nu} | \vect{d} ] $ is Gaussian, many efficient algorithms are available, and the algorithm actually used in the current study is specified in Supplement \ref{sec:app_sample} with,
\begin{align}
\vect{\nu}^s = [ \vect{\nu} | \vect{\nu} \in \mathcal{A} , \vect{d} ] &\rightarrow 
         [ \Phi_{n_r} ( \cup_{i=1}^{n_r} \mathcal{A}_i ;  \vect{\mu}_{\nu|d} , \matr{\Sigma}_{\nu | d} )]^{-1} 
              \times  \phi_{n_r} (\vect{\nu} ; \vect{\mu}_{\nu|d}, \matr{\Sigma}_{\nu |d} ) 
                   \times  I[ \vect{\nu} \in \cup_{i=1}^{n_r} \mathcal{A}_i ] \\
[ \vect{r}_A | \vect{d} ]^s = [ \vect{r} | \vect{\nu}^s , \vect{d} ] &\rightarrow  
  \phi_{n_r} ( \vect{r} ; \vect{\mu}_{r|\nu^s d}  , \matr{\Sigma}_{r|\nu^s d} ) \nonumber                  
\end{align}
where
\begin{align*}
\vect{\mu}_{r|\nu^s d} &= \vect{\mu}_{r|d} + \matr{\Gamma}_{r \nu | d} \matr{\Sigma}^{-1}_{\nu | d} 
                          [ \vect{\nu}^s - \vect{\mu}_{\nu | d} ] \\
\matr{\Sigma}_{r | \nu^s d} &= \matr{\Sigma}_{ r | d} - \matr{\Gamma}_{r \nu | d} \matr{\Sigma}^{-1}_{\nu | d} \matr{\Gamma}_{ \nu r| d}
\end{align*}.
Prediction of $ [ \vect{r}_A | \vect{d} ] $ need to be carefully designed since we often define selection Gaussian prior models with multiple modes, and so will also the posterior pdf be. 
The traditional expectation (E) predictor based on a minimum locationwise squared error loss, denoted 
$ \hat{\vect{r}}_E =\E \{ \vect{r}_A | \vect{d} \} $, will often appear in low-probability regions inbetween modes of the posterior pdf. The median (M) predictor based on a minimum locationwise absolute error criterion, denoted 
$ \hat{\vect{r}}_M =MED \{ \vect{r}_A | \vect{d} \} $, shear the same tendency to appear in low-probability regions. The prefered predictor is the global maximum posterior predictor, but it is usually too computer demanding to determine since it requires optimization of a $ n_r $-dimensional multi-modal function. Therefore we recommend the maximum posterior (MAP) predictor based on a maximum locationwise posterior criterion,
\begin{align}
\hat{\vect{r}}_{MAP} 
&= \mbox{MAP} \{ \vect{r}_A | \vect{d} \} \\
         & = \{ \mbox{MAP} \{ r_{Aj}| \vect{d} \} = \argmax_{r_j} \{f( r_j | \vect{d} ) \} ; j=1, \dots ,n_r \} \nonumber
\end{align}
with
\begin{align*}
  [r_{Ai} | \vect{d} ] = [ r_i | \vect{\nu} \in \mathcal{A}, \vect{d} ]  \rightarrow  &f( r_{Ai} | \vect{d} ) = f( r_i | \vect{\nu} \in \mathcal{A}, \vect{d} )  \\
                  &= [ F( \vect{\nu} \in \mathcal{A}| \vect{d} )]^{-1} F( \vect{\nu} \in \mathcal{A}| r_i, \vect{d} ) f( r_i | \vect{d} ) \\
                   &=[ \Phi_{n_r} ( \cup_{i=1}^{n_r} \mathcal{A}_i ; \vect{\mu}_{\nu|d} , \matr{\Sigma}_{\nu|d} )]^{-1} \\
                 &\times \Phi_{n_r} ( \cup_{i=1}^{n_r} \mathcal{A}_i ; \vect{\mu}_{\nu | r_i d},  \matr{\Sigma}_{\nu | r_i d} ) \\
                 &\times \phi_1 ( r_i; \mu_{r_i | d} , \sigma^2_{ r_i |d} )  
\end{align*}
with 
\begin{align*}
\left[
\begin{array}{c}
\mu_{r_i|d} \\  \vect{\mu}_{\nu|d}
\end{array}  \right]
= \left[
\begin{array}{c}
 \mu  \\ 0 \vect{i}_{n_r}
\end{array}  \right]
+ \left[
\begin{array}{c}
\sigma^2 \vect{c}_i^T \matr{H}^T  \\  \gamma \sigma \matr{C}  \matr{H}^T 
\end{array}  \right]
\left[
\sigma^2 \matr{H} \matr{C} \matr{H}^T + \matr{\Sigma}_{d|r}
\right]^{-1}
\left[
\vect{d} - \mu \matr{H}  \vect{i}_{n_r}
\right] 
\end{align*}
\begin{align*}
\left[
\begin{array}{cc}
\sigma_{r_i|d}^2  &  \vect{\gamma}_{r_i \nu|d} \\
\vect{\gamma}_{\nu r_i|d} & \matr{\Sigma}_{\nu | d}
\end{array} \right]
=  &\left[
\begin{array}{cc}
\sigma^2   &  \gamma  \sigma \vect{c}_i^T  \\
\gamma \sigma \vect{c}_i  &  \gamma^2 \matr{C} + [1 - \gamma^2 ] \matr{I}_{n_r}
\end{array}  \right] \\
&- \left[
\begin{array}{c}
\sigma^2 \vect{c}_i^T \matr{H}^T \\
\gamma \sigma \matr{C} \matr{H}^T 
\end{array} \right]
\left[
\sigma^2 \matr{H} \matr{C} \matr{H}^T + \matr{\Sigma}_{d|r}
\right]^{-1}
\left[
\begin{array}{cc}
\sigma^2 \matr{H} \vect{c}_i  &  \gamma \sigma \matr{H}  \matr{C}
\end{array}  \right]
\end{align*}
\begin{align*}
\vect{\mu}_{\nu | r_i d} &= \vect{\mu}_{\nu | d} + \vect{\gamma}_{vr_i|d} \sigma_{r_i|d}^{-1} [r_i - \mu_{r-i|d}] \\
\matr{\Sigma}_{\nu | r_i d} &= \matr{\Sigma}_{\nu |d} - \vect{\gamma}_{\nu r_i|d} \sigma_{r_i|d}^{-1} \vect{\gamma}_{r_i \nu | d}
\end{align*}
which normally appear close to the dominant mode of the posterior pdf. This predictor is relatively simple to identify since the marginal posterior pdfs are known to be selection Gaussian pdfs with analytically assessable parameter values. The associated prediction $ \alpha $-intervals will naturally be the interval between the upper/lower $ \alpha/2 $-quantiles of the  marginal posterior pdfs, which usually must be assessed by simulation based inference. Note also, that these predictors and prediction intervals will correspond to kriging if the prior pdf is from the pure Gaussian class since MAP and E predictors coinside for uni-modal symmetrical pdfs.

\begin{figure}
 \centering
\includegraphics[width=0.7\textwidth]{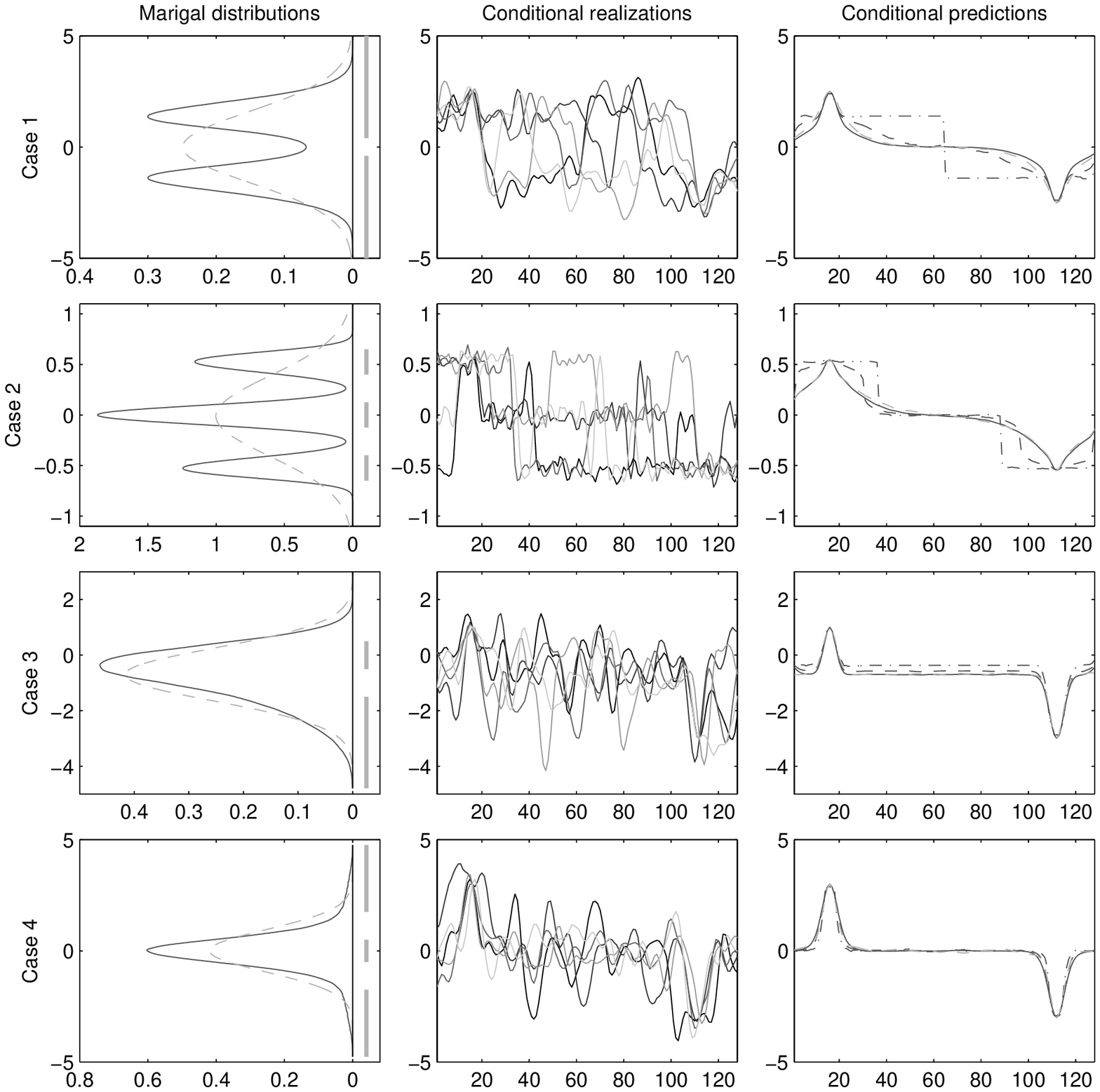}
 \caption{First column: marginal distribution of prior selection Gaussian model ( solid black) and corresponding Gaussian model ( dashed gray), and selection set on auxiliary random field on axis (solid gray). Second column: five realizations of the posterior selection Gaussian random field. Third column: posterior selection Gaussian model predictions, with  E-prediction ( solid black), MED-prediction ( dashed black), and MAP-prediction (dashed-dotted black). The corresponding Gaussian model prediction (E/MED/MAP)  ( dashed gray).}
 \label{fig:synthetic_prediction}
\end{figure}

The characteristics of the posterior selection Gaussian spatial model are exhibited in Figure \ref{fig:synthetic_prediction} which is based on the parameter sets listed in Table \ref{tbl:synthetic_param_prediction}. A more detailed discussion of the examples is presented in Supplement \ref{sec:app_exPosterior}. The spatial variable is represented on a $ 128$-grid with exact observations in grid nodes $ 16 $ and  $ 112 $. The different prior models produces very different posterior realizations and predictions, all of them exactly honoring the observations of course. The MAP-predictor is particularly sensitive to multi-modal marginals in the prior model. The computer demand for this simple example is very modest since the posterior model is analytically tractable.

\begin{table}
\caption{Model parameters for four posterior cases, with $\mu = 0 $ and  $\sigma^2=1$ for all cases.}
\begin{center}
\begin{small}
\begin{tabular}{c|ccccc }
Case & $\gamma$ & $d_h$ & $A_i$ & description & cond. values\\ \hline
1    & 0.900  & 4   & $(\infty, -0.4] \cup [0.4, \infty)$ & sym. bimodal & $ 2.5, -2.5$\\ 
2    & 0.999  & 4   & $[-0.65, -0.4] \cup [0.12, 0.12] \cup [0.4 0.65]$    & sym. trimodal & $ 0.55, -0.55$\\
3    & 0.600  & 4   & $(\infty, -1.5] \cup [-0.5, 0.5)$ & asym. unimodal & $1.0, -3.0$\\
4    & 0.700 & 4   & $(\infty, -1.75] \cup [-0.5, 0.5] \cup [1.75, \infty)$& sym. heavy tailed & $ 3.0, -3.0$\\
\end{tabular}
\end{small}
\end{center}
\label{tbl:synthetic_param_prediction}
\end{table}

\section{Model Parameter Inference}

One challenge with this class of selection Gaussian pdfs is the lack of clear interpretation of the model parameters, even in the reduced parametrization used as spatial stationary prior pdf in this study. The fact that the model parameter values are dependent on the actual grid-design $ \mathcal{L}_\mathcal{D} $ , complicates matters even more. In this section we discuss model parameter inference in some larger detail.

In order to perform Bayesian  inversion, all model parameters of both the likelihood function and the prior pdf must be assigned values. The likelihood parameters $ \vect{\theta}_l $ are assumed to be known  through studies of the observation acquisition procedure. The model parameters of the prior pdf $ \vect{\theta}_p $ are more complicated to elicit.

One may consider a hierarchical Bayesian inversion model, combining Bayesian inversion and Bayesian inference, 
and consider $ \vect{\theta}_p 
$ as a random variable with a suitable prior model $ f ( \vect{\theta}_p ) $. Then, in principle, the posterior model for $ \vect{\theta}_p $ is available,
\begin{align*}
[ \vect{\theta}_p | \vect{d} ] \rightarrow f ( \vect{\theta}_p | \vect{d} ; \vect{\theta}_l ) 
= \mbox{const} \times \int f( \vect{d} | \vect{r}_A ; \vect{\theta}_l ) f( \vect{r}_A | \vect{\theta}_p ) f ( \vect{\theta}_p ) d \vect{r}_A.
\end{align*}
Remember that $ \vect{\theta}_p = ( \mu, \sigma^2, \gamma, \rho( \vect{\tau}), \mathcal{A}_i) $ where $ \mu \in \mathcal{R} $ ,
 $ \sigma^2 \in \mathcal{R}_+ $ and $ \gamma \in [-1,1] \subset \mathcal{R} $ while $ \rho(0) = 1 $ and $ \rho( \vect{\tau}) ; \vect{\tau} \in \mathcal{R}_+^m $ is a positive definite function, and $ A_i \subset \mathcal{R} $. Hence, both assigning suitable prior models to $ \vect{\theta}_p $ and calculation of the normalizing constant appear as very complicated.

We recommend using training images of the spatial variable of interest and discretize them to a grid-design corresponding to $ \mathcal{L}_\mathcal{D} $, ie. the same grid spacing along all dimensions. These discretized training images will in many applications be available since the grid-design often is defined by the observation acquisition procedure and hence used in several sites. Denote one such discretized training image by $ n_r^o $-vector $ \vect{r}_A^o $. The corresponding Bayesian inference expression is,
\begin{align*}
[ \vect{\theta}_p | \vect{r}_A^o ] \rightarrow f ( \vect{\theta}_p | \vect{r}_A^o ) = \mbox{const} \times f ( \vect{r}_A^o | \vect{\theta}_p ) f ( \vect{\theta}_p )
\end{align*}
which also will be very complicated to assess for the full model parameter vector $ \vect{\theta}_p $. 
In \citet{Arellano-Valle2009} and \citet{Branco2013} this posterior model for the model parameter $ \gamma $ given the other parameters in $ \vect{\theta}_p $ and with $ \mathcal{A}_i = \mathcal{R}_\oplus $ is discussed. The authors of the former reference also provide guidelines for obtaining conjugate prior models for $ \gamma $. The generalization of these results to cover the full prior model parameter vector $ \vect{\theta}_p $ appears as very complicated.

We choose to continue in a classical inference setting and develop the log-likelihood for observing $ \vect{r}_A^o $ as a function of $ \vect{\theta}_p $,
\begin{align} 
\log L(  \vect{\theta}_p ,  \vect{r}_A^o ) &= \log f (  \vect{r}_A^o ; \vect{\theta}_p ) \\
         &= - \log  \Phi_{n_r^o} ( \cup_{i=1}^{n_r^o} \mathcal{A}_i ; 0 \vect{i}_{n_r^o} , \gamma^2 \matr{C} + [1-\gamma^2]               
                 \matr{I}_{n_r^o}) \nonumber \\
            &+ \sum_{i=1}^{n_r^o} \log \Phi_1 ( \mathcal{A}_i ; \gamma \sigma^{-1} [r_{Ai}^o - \mu], [1-\gamma^2]) \nonumber \\
            &+ \log \phi_{n_r^o} ( \vect{r}_A^o ; \mu \vect{i}_{n_r^o} , \sigma^2 \matr{C} ) \nonumber
\end{align}
In theory we may then define the maximum likelihood estimator for $ \vect{\theta}_p  $  by,
\begin{align} \label{eqn:param_likelihood}
 \hat{ \vect{\theta}_p } = \argmax_{\vect{\theta}_p} \{ \log L(  \vect{\theta}_p ,  \vect{r}_A^o )  \}
\end{align}
but in practice we need to parametrize also $ \rho( \vect{\tau} ) $ and $ A_i $ in order to perform the optimization, and this parametrization will be problem specific. One major challenge in the optimization is that the probability $ \Phi_{n_r^o} ( \cdot ) $ need to be recalculated for varying   $ \vect{\theta}_p  $, which may be extremely computer demanding. We calculate this probability by an importance blocking rejection algorithm, see  Supplement \ref{sec:app_is}. Lastly, there is no guarantee that the object function $  \log L(  \vect{\theta}_p ,  \vect{r}_A^o )  $ may not be multi-modal, which makes optimization notoriously complicated. It will be unfair to say that not many unresolved issues remain, but we present one encouraging example of prior model parameter elicitation based on training images below.

\begin{figure}
 \centering
\includegraphics[width=0.55\textwidth]{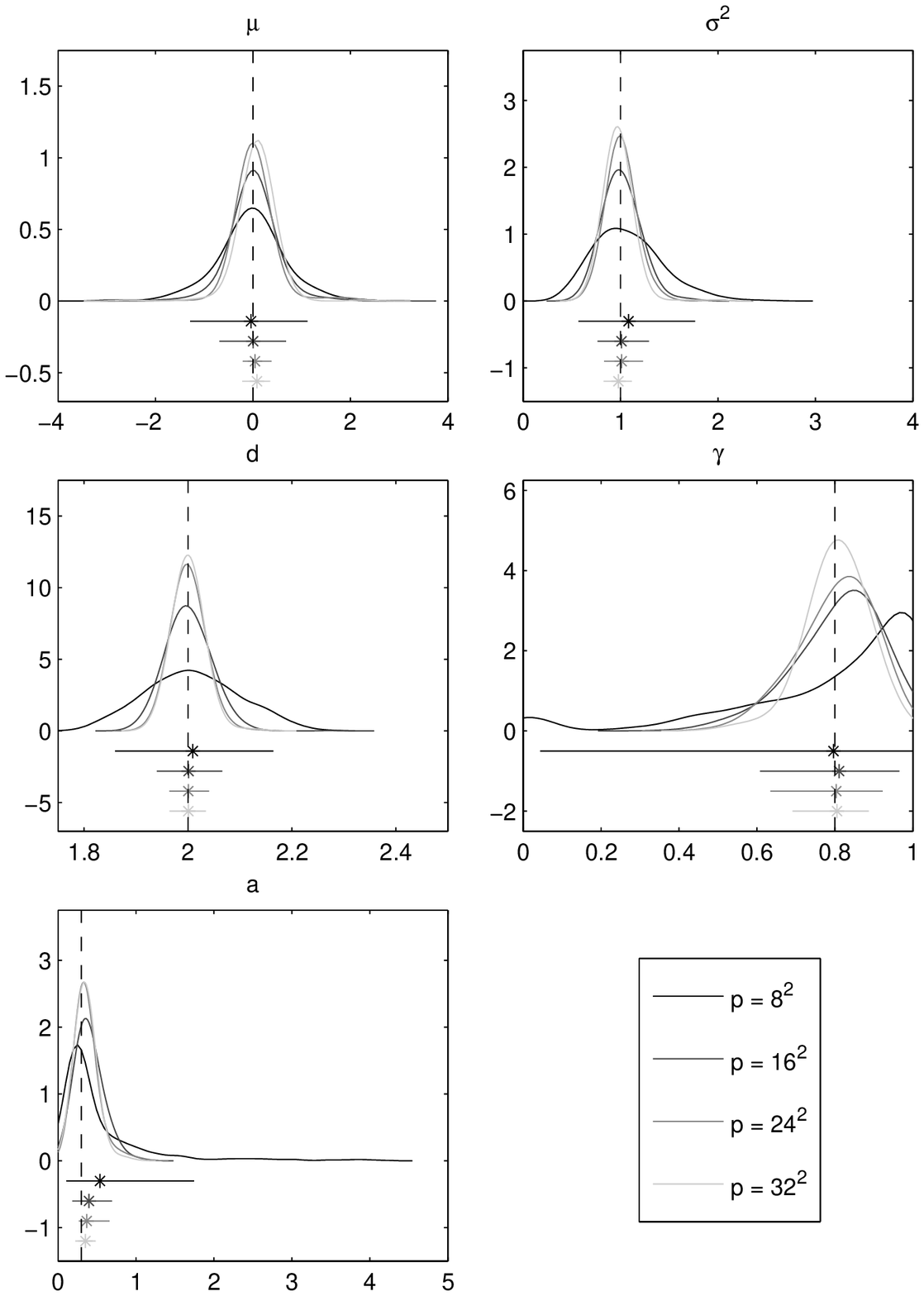}
 \caption{Density plots of parameter estimates $\hat{ \vect{\theta}}_p $ with increasing size of the training image $ \vect{r}^o $. Below are means and 90\% confidence intervals, and true values ( vertical dashed lines).}
 \label{fig:para_est}
\end{figure}

We evaluate the characteristics of the maximum likelihood estimator for $ \vect{\theta}_p = ( \mu, \sigma^2, d, \gamma, a) $, where $ d = d_h = d_v$ and $ \mathcal{A}_i : (- \infty, -a] \cup [a, \infty) $ , for case 1 in Figure \ref{fig:synthetic_marginals}  and Table \ref{tbl:synthetic_param}. The results are exhibited in Figure \ref{fig:para_est}, and a more detailed discussion is presented in Supplement \ref{sec:app_exInference}. The training images are subsets of the realization in Figure \ref{fig:synthetic_marginals} of sizes $ [8 \times 8] , [16 \times 16], [24 \times 24] $ and $ [32 \times 32] $. By repeating this inference on $ 1000 $ realizations from the prior model we can assess the accuracy and precision of the estimator. We observe from Figure \ref{fig:para_est} that the estimator appears as biased, but consistent as the training image increases, which is as expected for maximum likelihood estimators for ergodic spatial models. Moreover, it appears as relatively reliable estimates can be obtained even for training images of size $ [24 \times 24] $. The computer demand for estimating $ \vect{\theta}_p $ for one training image of size $ [32 \times 32] $ is typically one minute on a regular laptop computer.

\section{Case Study - Seismic inversion}

The objective of seismic inversion is to predict the elastic material properties - pressure-wave velocity, shear-wave velocity and density - in the subsurface based on observed amplitude-versus-offset (AVO) seismic data collected at the surface. The data appear as time-laged, angle-dependent reflection intensities from the subsurface created by an air pulse generated at the surface. We model the log-transformed properties in order to have a linear likelihood function, see \citet{buland:185}, $ \vect{r} = ( \log \vect{v}_p, \log \vect{v}_s, \log \vect{\rho} ) \in \mathcal{R}^{3n_r} $ and $ \vect{d} = ( \vect{d}^1, \vect{d}^2, \vect{d}^3 ) \in \mathcal{R}^{3n_d} $ with upper-index representing three angles. Hence the objective is to assess $ [ \vect{r} | \vect{d} ] $, and we phrase the inversion in a Bayesian setting.

\begin{figure}
\centering
\begin{subfigure}{.5\textwidth}
  \centering
  \includegraphics[width=.4\linewidth]{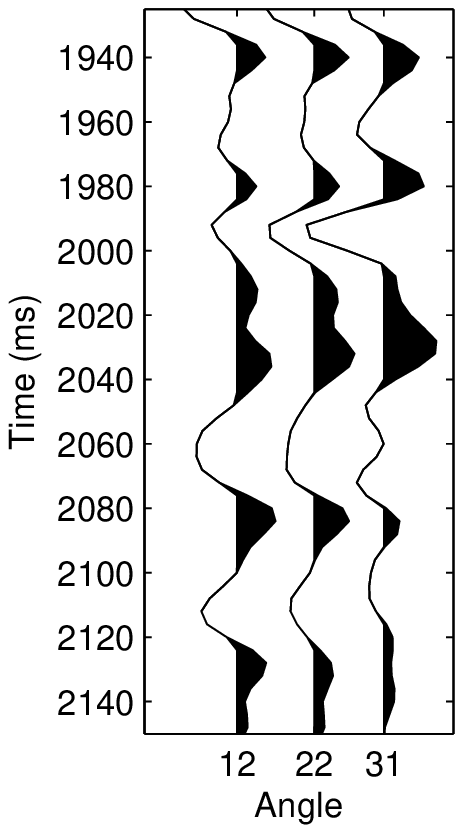}
  \caption{}
  \label{fig:alvheim_seismic}
\end{subfigure}%
\begin{subfigure}{.5\textwidth}
  \centering
  \includegraphics[width=.4\linewidth]{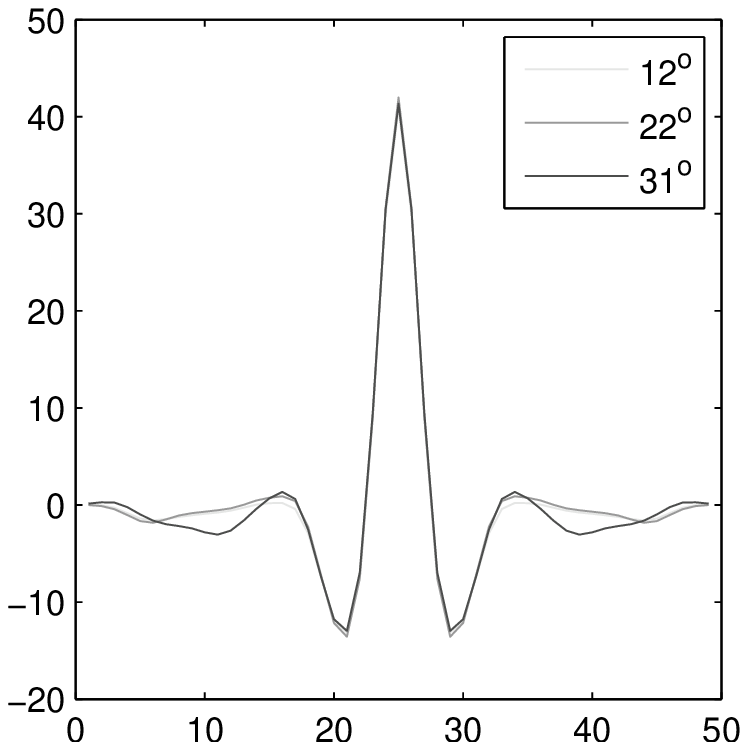}
  \caption{}
  \label{fig:alvheim_wavelets}
\end{subfigure}
\caption{Seismic AVO data in the well trace for reflection angles $12 \ensuremath{^\circ}$, $22 \ensuremath{^\circ}$, and $31 \ensuremath{^\circ}$, with depth in seismic two-way traveltime (a) and Seismic wavelets shape for reflection angles $12 \ensuremath{^\circ}$, $22 \ensuremath{^\circ}$, and $31 \ensuremath{^\circ}$ (b).}
\label{fig:test} 
\end{figure}

\begin{figure}
 \centering
\includegraphics[width=0.5\textwidth]{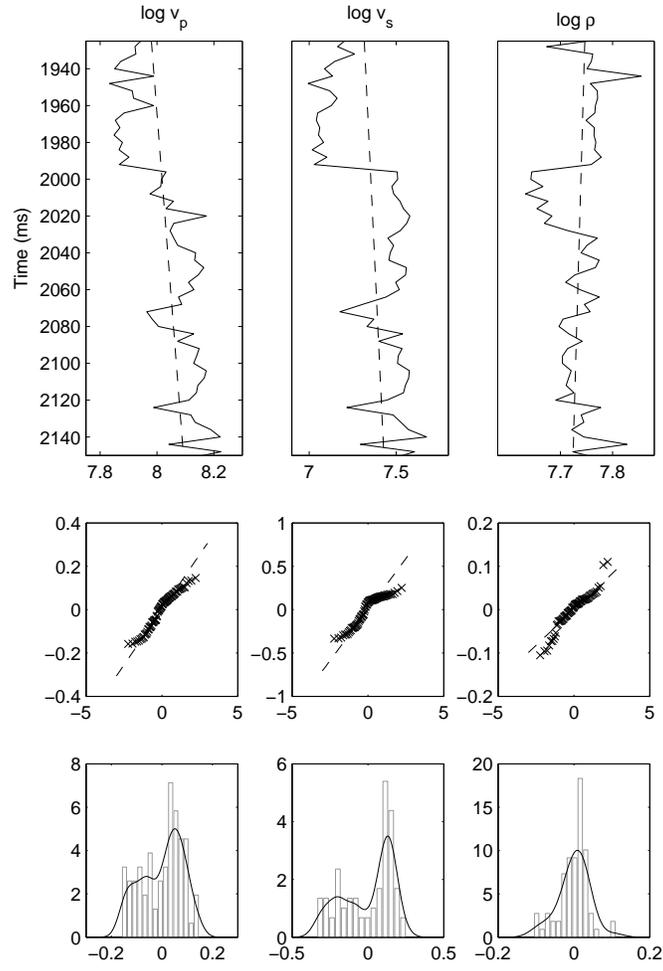}
 \caption{Well observations of logarithm of pressure-wave velocity $v_p$, share-wave velocity $v_s$, and density $\rho$. Top: elastic properties in the well with estimated linear trend ( dashed black). Middle: quantile-quantile plot of residual elastic properties. Bottom: histograms and density estimates of residual elastic properties.}
 \label{fig:alvheim_well}
\end{figure}

The case study is based on data from the Alvheim field in the North Sea ( \citet{avseth:788} , \citet{Rimstad2011}). The subsurface contains a turbiditic oil and gas reservoir at about 2000 meters depth, but we use reflection time as depth reference with one meter (m) corresponding to approximately one milli-second (ms). We consider one vertical profile at the depth range $ \mathcal{D} : [ 1935-2145] $ ms discretized to $ \mathcal{L}_\mathcal{D} $ with $ n=n_r = n_d = 55 $ grid nodes, where both AVO seismic data $ \vect{d} $ and exact observations of the elastic material properties $ \vect{r}^o $ in a well, are available, see Figure \ref{fig:alvheim_seismic}  and  \ref{fig:alvheim_well}. Both $ \vect{d} $ and $ \vect{r}^o $ are used to infer the likelihood and prior model parameters $ \vect{\theta}_l $ and $ \vect{\theta}_p $, by considering $ [ \vect{d} | \vect{r}^o ]$ and $ \vect{r}^o $ as training images, respectively. In the Bayesian spatial inversion we consider $ [ \vect{r} | \vect{d} ] $, hence the posterior model is only conditioned on $ \vect{d} $. The training image  $ \vect{r}^o $ is used to validate the results. 
In practical use, seismic inversion of profiles in the neighborhood of the well trace, without well observations, will be made, but then model validation is complicated.
We perform Bayesian inversion based on two alternative prior models, one selection Gaussian and one traditional Gaussian prior model, and we compare the corresponding posterior models.

The likelihood model $ f( \vect{d} | \vect{r} ) $ link the seismic data $ \vect{d} $ and the elastic material properties of interest $ \vect{r} $. The model is based on a linearization of the wave equation as defined in \citet{buland:185},
\begin{align}
[ \vect{d} | \vect{r} ] &= \matr{W} \matr{A} \matr{D} \vect{r} + \vect{\epsilon}_{d|r} \\
 &\rightarrow f ( \vect{d} | \vect{r } ) = \phi_{3n_r} ( \vect{d} ; \matr{W} \matr{A} \matr{D} \vect{r} , \sigma^2_{d|r} \matr{\Sigma}_{d|r}^o ) \nonumber
\end{align}
where $ \matr{W} $ is a convolution matrix defined by the kernels in Figure \ref{fig:alvheim_wavelets}; matrix $ \matr{A}$ represents the angle-dependent linearized wave equation; $ \matr{D} $ is a diffentiation matrix; and $ \vect{\epsilon}_{d|r} $ is a centred Gaussian vector with covariance matrix $ \matr{\Sigma}_{d|r} = \sigma^2_{d|r} \matr{\Sigma}_{d|r}^o $. The correlation matrix $ \matr{\Sigma}_{d|r}^o $ is defined by exponential correlation functions in angle and time with ranges $ d^a $ and $ d^t $, respectively. Hence the likelihood model is Gauss-linear with parameters $ \vect{ \theta}_l = ( \sigma^2_{d|r} , d^a , d^t ) $.

The prior model $ f( \vect{r} ) $ represents the general characteristics of the elastic material properties of interest. Figure \ref{fig:alvheim_well}  contain a plot of the exact observations of the properties $ \vect{r}^o $ along the profile plus Gaussian quantile-quantile plots and histograms of residuals after removing the linear vertical trend. The bi-modality of the histograms of $ \log v_p $ and $ \log v_s $ are caused by vertically varying rock types in the subsurface. By using a selection Gaussian prior model, this bi-modality in the marginal pdfs can be captured. The model as defined in Section  2.3  must be extended to represent the tri-variate $ \vect{r} $, and the parametrization become 
$ \vect{\theta}_p = ( \vect{\mu}, \matr{\Sigma} , \vect{\gamma}, d^r, \vect{a} ) $. The spatial exponential correlation function with range $ d^r $ is common for all three variables and the selection sets are parametrized as 
$ \mathcal{A}_i : \{ ( - \infty, a] , [a, \infty ) \} $ with specific $ a $ values for each variable. The alternative traditional Gaussian prior model $ f^G ( \vect{r} ) $ has parametrization $ \vect{\theta}^G_p =( \vect{\mu}, \matr{\Sigma}, d^r ) $ and will not capture the bi-modality in the variables.

\begin{figure}
 \centering
\includegraphics[width=0.6\textwidth]{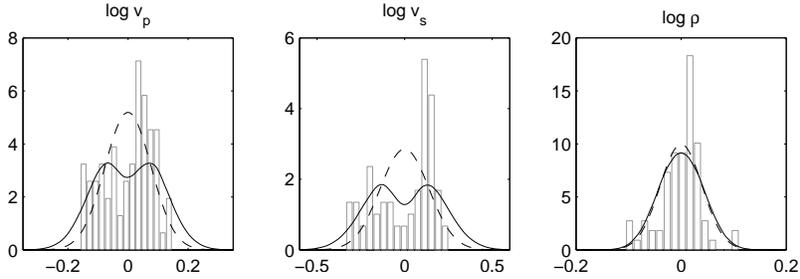}
 \caption{Estimated prior marginal models. Marginal distributions of estimated selection Gaussian random field ( solid black), marginal distributions of estimated Gaussian random field ( dashed black), and histograms of well observations.}
 \label{fig:alvheim_prior_select}
\end{figure}

We infer the likelihood parameters $ \vect{\theta}_l $ from the available seismic data $ \vect{d} $ and the exact observations of the elastic material properties $ \vect{r}^o $ by using a maximum likelihood criterium,
\begin{align*}
\hat{\vect{\theta}}_l = \argmax_{\vect{\theta}_l} \{ p( \vect{d} | \vect{r}^o ; \vect{\theta}_l) \}
\end{align*}
Likewise we infer the model parameters for the two alternative prior models, $ \vect{\theta}_p $ and $ \vect{\theta}_p^G $ from $ \vect{r}^o $. We set the location parameter $ \vect{\mu} $, for both models, equal to the vertical linear trend for each of the three variables, and estimate the remaining parameters by a maximum likelihood criterium,
\begin{align*}
\hat{\vect{\theta}}_p = \argmax_{\vect{\theta}_p | \vect{\mu} - \mbox{trend} \, \vect{r}^o} 
\{ f( \vect{r}^o ; \vect{\theta}_p ) \}  \\
\hat{\vect{\theta}}_p^G = \argmax_{\vect{\theta}_p^G | \vect{\mu} - \mbox{trend} \, \vect{r}^o} 
\{ f^G( \vect{r}^o ; \vect{\theta}_p^G ) \}
\end{align*}
The optimizations of the likelihood functions all appear to converge to unique optima with computer demands of a few minutes on a regular lap-top computer. The actual estimates for the likelihood model parameters are,
\begin{align*}
\hat{\vect{\theta}}_l:  \sigma^2_{d|r} = 0.402, d^a = 7.3, & \; d^t = 11.1,  \notag \\
\end{align*}
while the estimates for the two alternative prior model parameters are,
\begin{align*}
\hat{\vect{\theta}}_p: 
&\matr{\Sigma}
=
\left[
\begin{array}{rrr}
    0.0073  &  0.0126 &  -0.0013 \\
    0.0126  &  0.0250 &  -0.0039 \\
   -0.0013  & -0.0039 &   0.0018
\end{array}
\right], \\ 
&\vect{\gamma} = \left[
\begin{array}{r}
  0.8656 \\   0.9061 \\   0.3331
\end{array}
\right],
d^r = 1.61,
\vect{a} = \left[
\begin{array}{r}
    0.1110 \\   0.2619 \\   0.1151
\end{array}
\right]
.
\end{align*}
\begin{align*}
\hat{\vect{\theta}}_p^G: 
\matr{\Sigma}
=
\left[
\begin{array}{rrr}
    0.0059 &   0.0093 &  -0.0007 \\
    0.0093 &   0.0195 &  -0.0025 \\
   -0.0007 &  -0.0025 &   0.0016
\end{array}
\right],
d^r = 1.53.
\end{align*}
The parameter estimates for the two alternative prior models appear as consistent with comparable range lengths and dependence structures between the three variables.  In Figure \ref{fig:alvheim_prior_select} the marginal pdfs of the two prior models are displayed together with the histograms of $ \vect{r}^o $. The selection Gaussian prior model captures the bi-modality of the histograms, without overfitting to the available well observations.

Based on the Gauss-linear likelihood model $ f( \vect{d} | \vect{r} ) $ with parameter values $ \hat{\vect{\theta}}_l $ and the selection Gaussian prior model $ f ( \vect{r} ) $ with parameter values $ \hat{\vect{\theta}}_p $, we use Bayesian spatial inversion to assess the posterior model $ f ( \vect{r} | \vect{d} ) $ which also will be selection Gaussian.
By using the alternative traditional Gaussian prior model with associated parameter values we obtain a Gaussian posterior model $ f^G ( \vect{r} | \vect{d} ) $.

\begin{figure}
\centering
\includegraphics[width=0.4\textwidth]{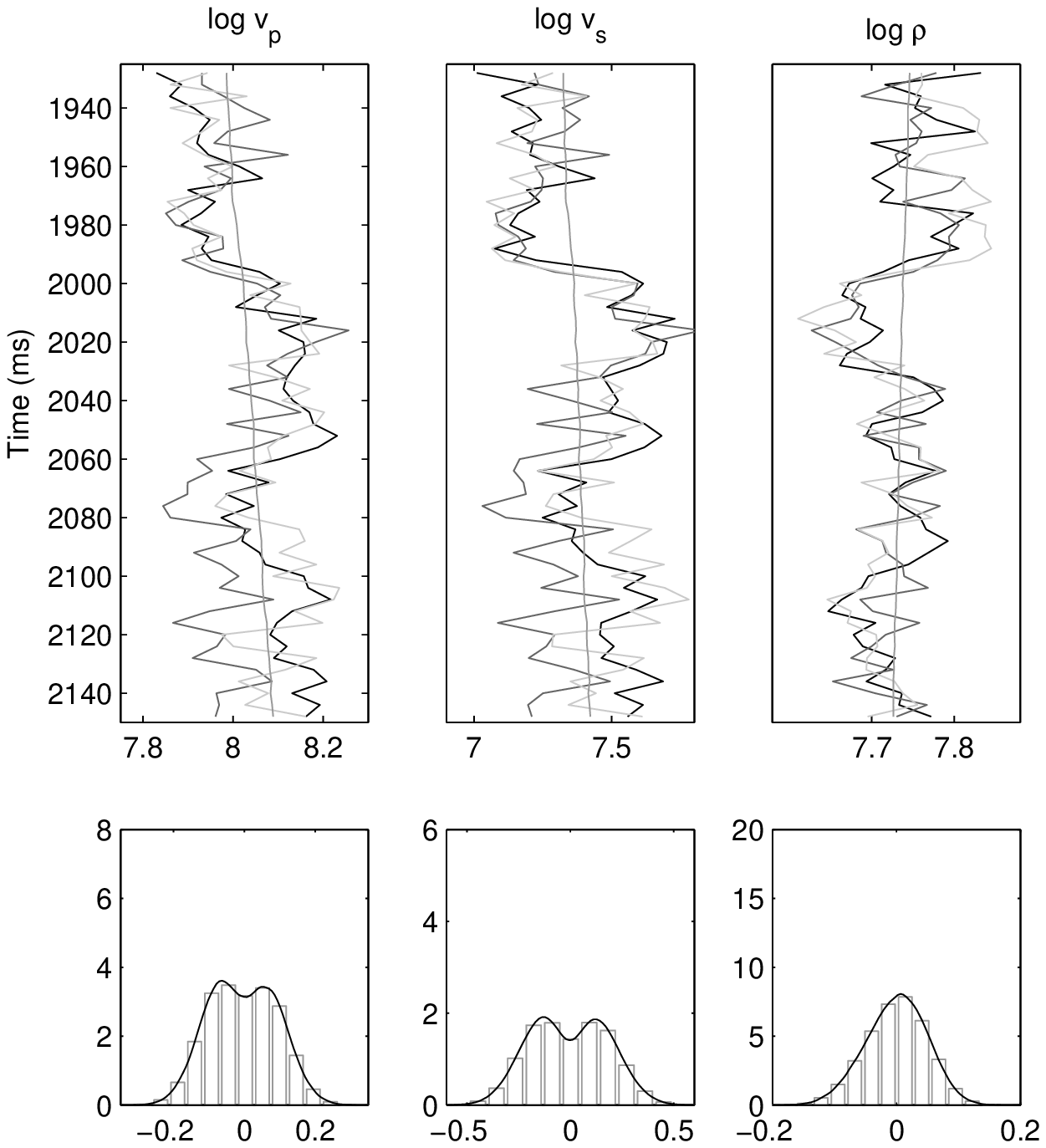}
\includegraphics[width=0.4\textwidth]{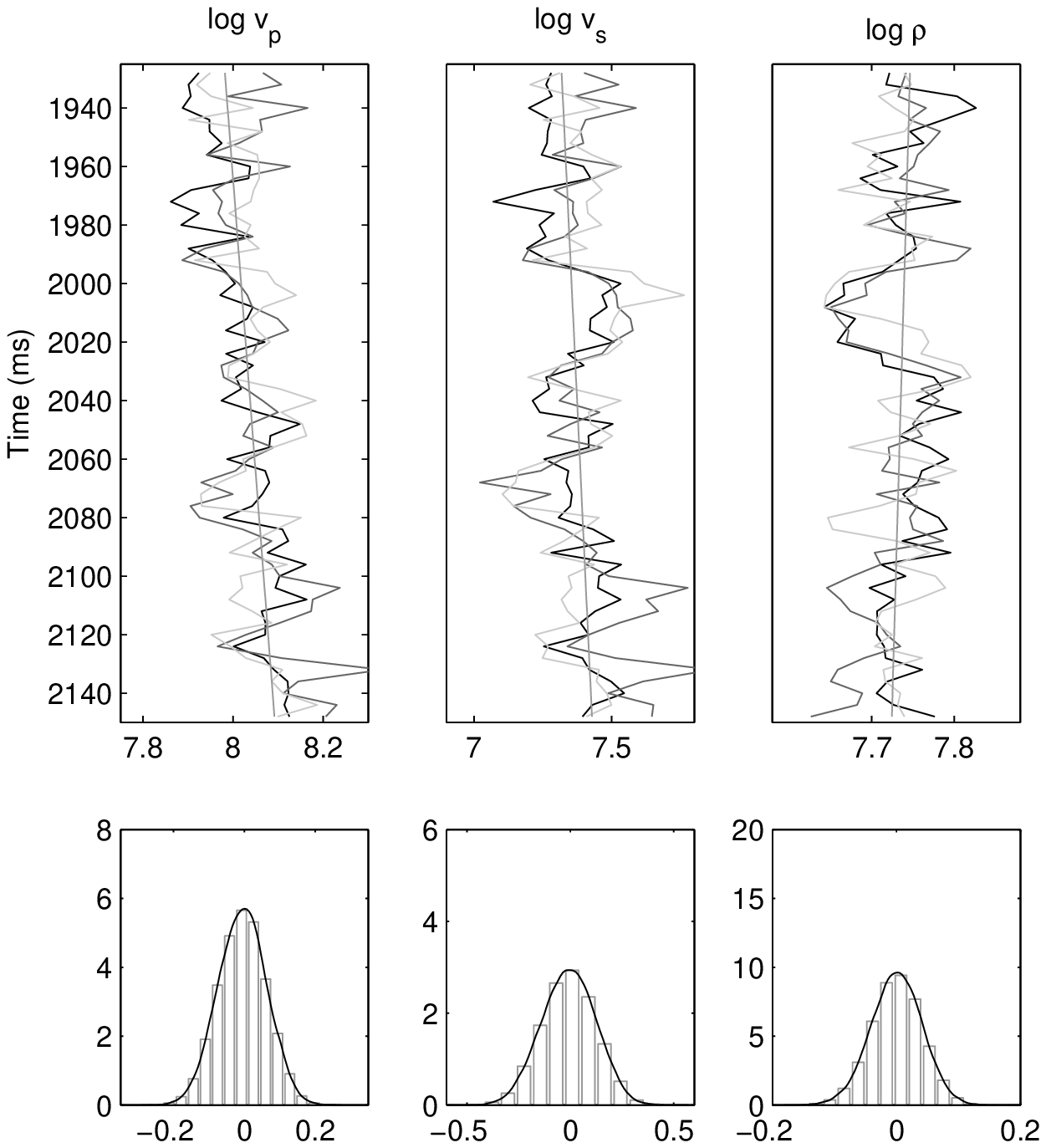}
 \caption{Three simulated realizations from posterior random fields, and realizations integrated over time. Top: Selection Gaussian model. Bottom: Traditional Gaussian model.}
 \label{fig:alvheim_pred_real}
\end{figure}

\begin{figure}
\centering
\includegraphics[width=0.4\textwidth]{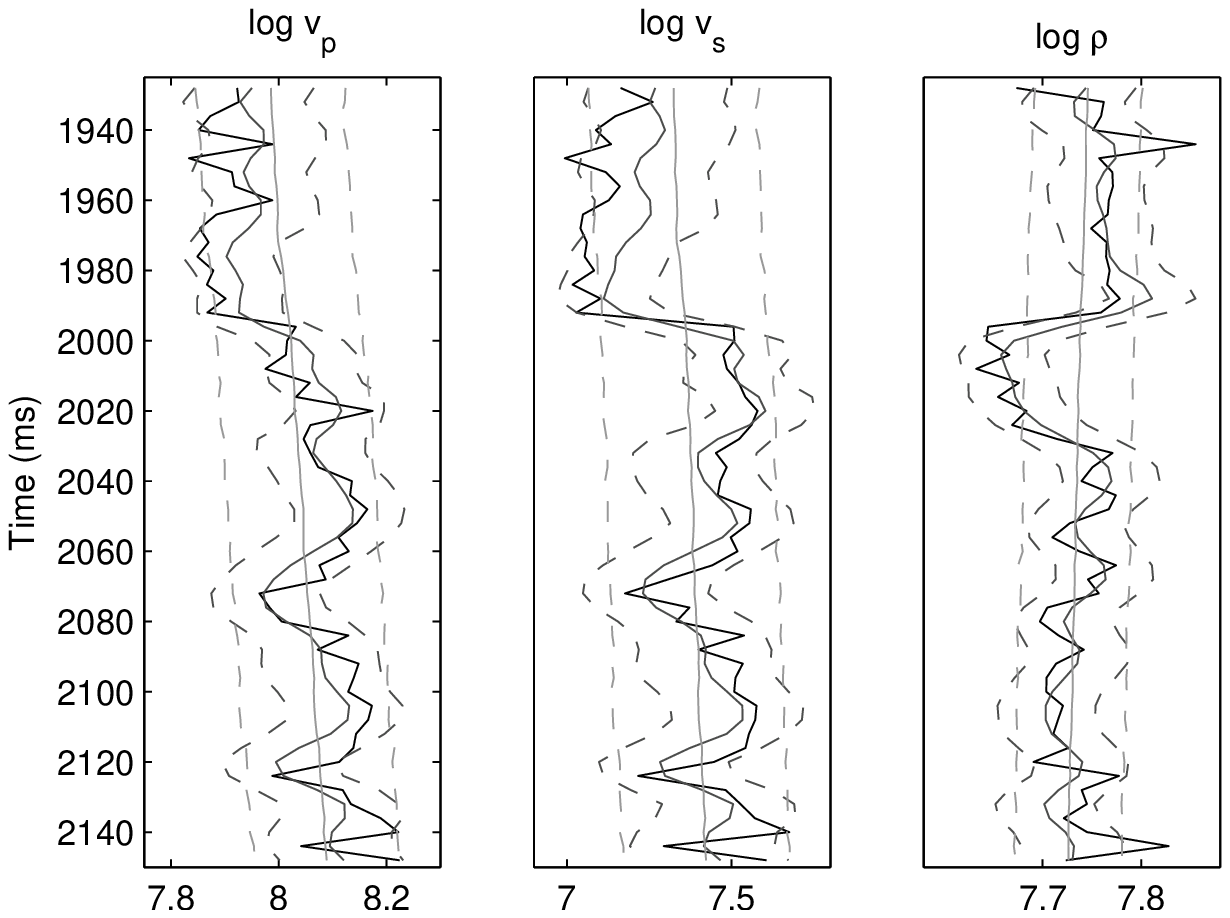}
\includegraphics[width=0.4\textwidth]{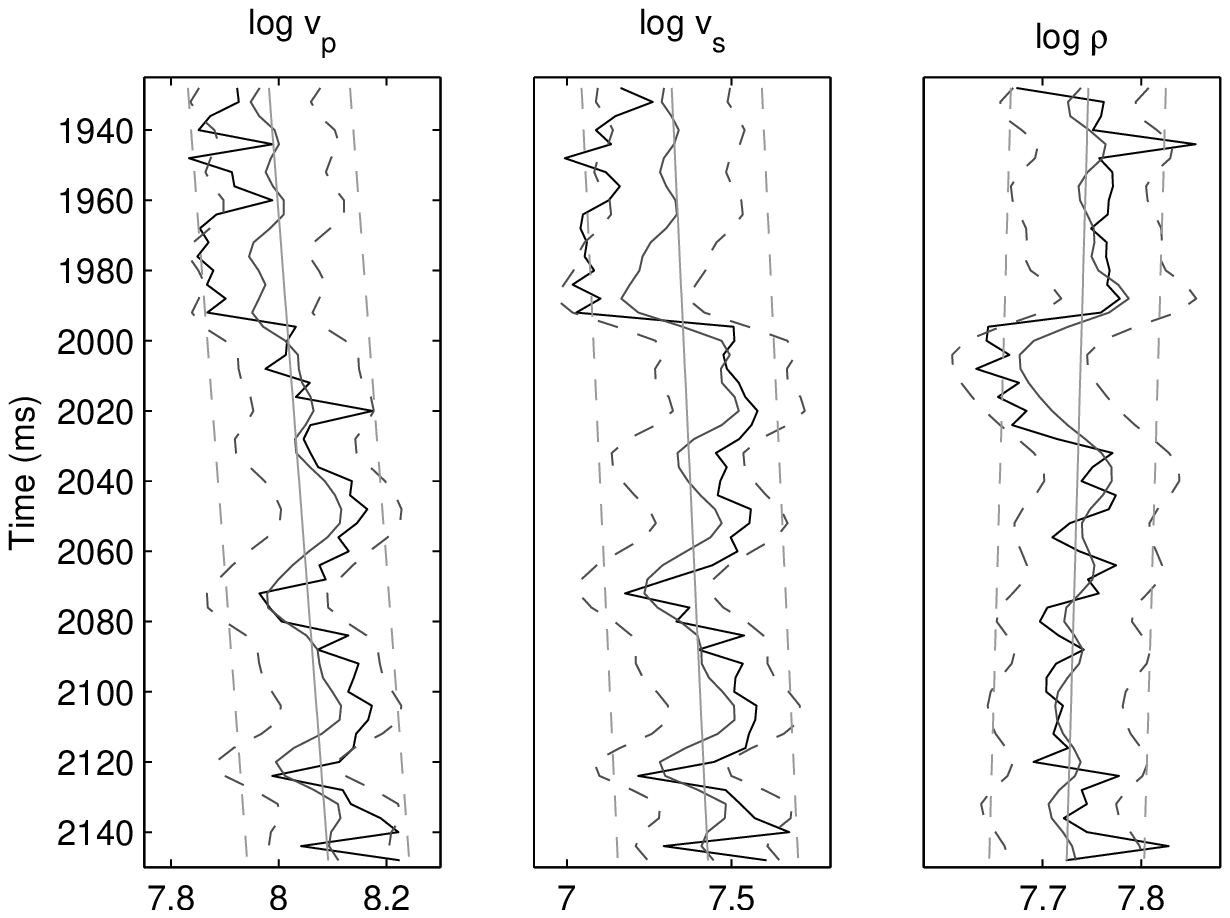}
 \caption{Well predictions.  Top: selection Gaussian model. Bottom:Traditional  Gaussian model.  Well observations ( solid black), posterior mean ( solid dark gray), posterior 80\% prediction interval ( dashed dark gray), prior mean ( solid light gray), and prior 80\% prediction interval ( dashed light gray).}
 \label{fig:alvheim_pred}
\end{figure}

Realizations from the two alternative posterior models,  $ f ( \vect{r} | \vect{d} ) $ and  $ f^G ( \vect{r} | \vect{d} ) $ are displayed in Figure \ref{fig:alvheim_pred_real}. The realizations from the former can be generated sequentially as outlined in Section \ref{Post_Mod}.  The realizations from the selection Gaussian posterior model appear with abrupt changes between two levels defined by the two modes in the prior model, hence the time-integrated histograms are bi-modal. The realizations from the Gaussian posterior model are smoother and the corresponding histograms are uni-modal.

Predictions of $ [ \vect{r} | \vect{d} ] $ based on the two alternative posterior models with associated 0.8-prediction intervals are displayed in Figure \ref{fig:alvheim_pred}. Also the correct elastic material property profiles $ \vect{r}^o $ are presented. Moreover, predictions and 0.8-prediction intervals for the two alternative prior models are displayed. We use E-predictors for both models. In this case study, contrary to the example in Section \ref{Post_Mod}, we have densely sampled data. The data consists of convolved, gradient observations in every node of $ \mathcal{L}_\mathcal{D} $, and the marginal pdf of the selection Gaussian posterior model will appear as almost uni-modal, although flipping between modes vertically. Consequently the E, MED and MAP predictors will be almost identical and the former is used for computational convenience. The predictions based on both posterior models do reproduce the correct profiles relatively well with large improvements of the prior predictions. The predictions from the selection Gaussian posterior model appear with more abrupt changes than the Gaussian one, whenever the correct profiles has large steps. Moreover, the 0.8-prediction intervals are narrower for the former model than for the latter.

\begin{table}
\caption{Summary of well predictions for the selection Gaussian and Gaussian model. Mean square error (MSE) of predictions, posterior and prior 80\% coverage of prediction intervals.}
\begin{center}
\begin{tabular}{|c|rr|rr|rr|}
\hline
& \multicolumn{2}{|c|}{MSE} & \multicolumn{2}{|c|}{Prior 80\% coverage} & \multicolumn{2}{|c|}{Posterior 80\% coverage} \\
	& Selection  	& Gaussian 	& Selection	& Gaussian 	& Selection & Gaussian 	\\
\hline
$\log v_p$	& 0.0034 	& 0.0050 	& 0.84		& 0.88 		& 0.85 & 0.96		\\
$\log v_s$	& 0.0112 	& 0.0191 	& 0.82		& 0.89 		& 0.84 & 0.87		\\
$\log \rho$	& 0.0009 	& 0.0011 	& 0.82		& 0.95		& 0.83 & 0.89		\\
\hline
\end{tabular}
\end{center}
\label{tbl:well}
\end{table}

Table \ref{tbl:well} contains summary statistics for Figure \ref{fig:alvheim_pred}. The mean square error (MSE) of the predictions relative to the correct profiles for each variable for both alternative models are listed.  Moreover, the coverage of the 0.8-prediction intervals for the correct profiles for the prior and posterior models are specified. The predictions from the selection Gaussian model appear as clearly superior to the predictions from the Gaussian one. The improvements in MSE are in the range of $ 20-40 $ \%. The coverage values for the selection Gaussian model is close to $ 0.8 $ as they should be, while the coverages for the Gaussian model are far too large and more variable.

\section{Concluding remarks}

We study Bayesian spatial inversion and introduce the consept of conjugate classes of prior parametrized pdfs with respect to given classes of likelihood functions. For this class of prior pdfs the associated posterior pdfs will be in the same class. Such conjugate classes exist for continuous, event and mosaic spatial variables for frequently used likelihood functions. The conjugate class of prior models can be selection extended without loss of the conjugate characteristic.

We demonstrate the potential of the selection extension by introducing the class of selection Gaussian prior pdfs which is conjugate with respect to Gauss-linear likelihood functions. The flexibility of this selection Gaussian class is displayed in a variety of examples which represent multi-modality, skewness and heavy-tailedness in the marginal distributions.

By using a prior model from a conjugate class for a given likelihood function, the associated posterior model can be assessed exactly based only on the model parameters of the prior and likelihood models - and the actual observations, of course. The normalizing constant, which usually complicates Bayesian spatial inversion, will be available on parametric form. We demonstrate this favorable characteristic for the class of selection Gaussian prior pdfs, and show that the posterior selection Gaussian pdf is analytically tractable which is used to make efficient algorithms for simulation and prediction. Several examples presenting conditional simulations and predictions exposing multi-modality, skewness and heavy-tailedness are displayed.

The class of selection Gaussian prior pdfs is parametrized by a number of model parameters which are not easily interpretable. Based on one training image of the spatial variable we define maximum likelihood estimators for the model parameters. A limited simulation study is conducted and we conclude that the estimators appear as consistent and that even for relatively small training images reliable estimates can be obtained. These results are encouraging.

Lastly, a case study using the selection Gaussian prior model on seismic inversion of real data is presented. We demonstrate $ 20$-$40\% $ improvement in the mean-square-error of predictions compared to the traditional Gaussian inversion. Also the prediction intervals of the former model appear as more reliable than for the latter.

The selection extension of conjugate classes of prior pdfs in Bayesian spatial inversion appears to have a large potential. We have to some extent explored this potential for continuous spatial variables and the class of selection Gaussian prior pdfs. The challenge for these models appears in sampling from and calculation of sub-set probabilities in high-dimensional Gaussian pdfs. We have presented some relatively efficient algorithms for these purposes. Many improvements of these algorithms are definitely possible. For event and mosaic spatial variables, the class of Poisson and Markov pdfs are conjugate with respect to certain likelihood functions. Also these classes can be selection extended and still remain conjugate. We have not yet explored these possibilities.

\section*{Acknowledgments}
The research is a part of the Uncertainty in Reservoir Evaluation (URE) activity at the Norwegian University of Science and Technology (NTNU).

\bibliographystyle{agsm}


\bibliography{ref}

\newpage

\appendix

\vspace{1cm} 

\appendix

\noindent
\textbf{\Large Appendix}

\section{Selection Gaussian Model} \label{app:selGauss}

The closedness properties of the selection Gaussian model is demonstrated.

\begin{Def}[Selection Gaussian pdf] \label{def:SGpdf}
Consider the $n$-vector  Gaussian basis-pdf,
\begin{align*}
\vect{r} \rightarrow f( \vect{r} ) = \phi_n ( \vect{r} ; \vect{\mu}_r , \matr{\Sigma}_r )
\end{align*}
and Gauss-linear  auxiliary $q$-vector variable,
\begin{align*}
[ \vect{\nu} | \vect{r} ] \rightarrow f ( \vect{\nu} | \vect{r} ) 
                 = \phi_q ( \vect{\nu} ; \vect{\mu}_{\nu|r} , \matr{\Sigma}_{\nu|r} )
\end{align*}
with $ \vect{\mu}_{\nu|r} = \vect{\mu}_\nu + \matr{\Gamma}_{\nu|r} ( \vect{r} - \vect{\mu}_r ) $, where 
$ \matr{\Gamma}_{\nu|r} $ is denoted the coupling $(q \times n)$-matrix.

Define a selection set $ \mathcal{A}_\nu \subset \mathcal{R}^q $, and the corresponding $n$-vector selection Gaussian pdf,
\begin{align*}
\vect{r}_A &= [ \vect{r} | \vect{\nu} \in \mathcal{A}_\nu ] 
                                     \rightarrow f(\vect{r}_A) = f ( \vect{r} | \vect{\nu} \in \mathcal{A}_\nu ) \\
               &= [ \Phi_q ( \mathcal{A}_\nu ; \vect{\mu}_\nu , \matr{ \Sigma}_\nu ) ]^{-1}
               \times \Phi_q ( \mathcal{A}_\nu ; \vect{\mu}_{\nu|r} , \matr{\Sigma}_{\nu|r} ) 
               \times \phi_n ( \vect{r} ; \vect{\mu}_r , \matr{\Sigma}_r )  \nonumber \\
               &= \mbox{const}
               \times \Phi_q ( \mathcal{A}_\nu ; \vect{\mu}_{\nu|r} , \matr{\Sigma}_{\nu|r} ) 
               \times \phi_n ( \vect{r} ; \vect{\mu}_r , \matr{\Sigma}_r )  \nonumber 
\end{align*}
with the covariance $ (q \times q) $-matrix 
$ \matr{\Sigma}_\nu = \matr{\Gamma}_{\nu|r} \matr{\Sigma}_r \matr{\Gamma}_{\nu|r}^T + \matr{\Sigma}_{\nu|r} $.

The class of selection Gaussian pdfs is defined by all valid sets of parameters  
$ ( \vect{\mu}_r , \matr{\Sigma}_r , \vect{\mu}_\nu, \matr{\Gamma}_{\nu|r}, \matr{\Sigma}_{\nu|r}, \mathcal{A}_\nu ) $.

\end{Def}

\noindent
The following results are useful for later Proofs,

\begin{Res}[Conditional Probabilities]
Consider the joint $ (n+m)$-vectorial variable $ (\vect{x} , \vect{y} ) $ with joint pdf $ f ( \vect{x}, \vect{y} ) $, then,
\begin{align*}
R1. \mbox{     }  f ( \vect{x} ) = \int_{\Omega_y} f ( \vect{x}, \vect{y} ) d \vect{y} 
              = \int_{\Omega_y} f ( \vect{x}| \vect{y} ) f( \vect{y} )  d \vect{y}
              =  \E_y \{ f ( \vect{x}| \vect{y} ) \}
\end{align*}
and also for arbitrary subset $ \mathcal{A}_x \subset \Omega_x $,
\begin{align*}
R2. \mbox{     }  F(\vect{x} \in \mathcal{A}_x ) = \int_{\mathcal{A}_x } f ( \vect{x} ) d \vect{x}
                      =  \int_{\mathcal{A}_x }  \E_y \{ f ( \vect{x}| \vect{y} ) \} d \vect{x}
                      = \E_y \{ F ( \vect{x} \in \mathcal{A}_x | \vect{y} ) \}
\end{align*}
For $ f ( \vect{x} , \vect{y} ) $ being a Gaussian pdf,
\begin{align*}
&R1G. \mbox{     }  \phi_n ( \vect{x} ; \vect{\mu}_x , \matr{\Sigma}_x )  
                        = \E_y \{ \phi_n ( \vect{x} ; \vect{\mu}_{x|y} , \matr{\Sigma}_{x|y} ) \}  \\
&R2G. \mbox{     }  \Phi_n ( \mathcal{A}_x ; \vect{\mu}_x , \matr{\Sigma}_x )  
                        = \E_y \{ \Phi_n ( \mathcal{A}_x ; \vect{\mu}_{x|y} , \matr{\Sigma}_{x|y} ) \}
\end{align*}
\end{Res}

The major statements are captured in the following Proposition,

\begin{Prop}[Selection Gaussian Models]
Consider the selection Gaussian prior model,
\begin{align*}
\vect{r}_A \rightarrow f ( \vect{r}_A ) = f ( \vect{r} | \vect{\nu} \in \mathcal{A}_\nu )
       = \mbox{const}  \times \Phi_q ( \mathcal{A}_\nu ; \vect{\mu}_{\nu|r} , \matr{\Sigma}_{\nu|r} ) 
               \times \phi_n ( \vect{r} ; \vect{\mu}_r , \matr{\Sigma}_r )
\end{align*}
and Gauss-linear $m$-vector likelihood model,
\begin{align*}
[ \vect{d} | \vect{r}_A ] \rightarrow f ( \vect{d} | \vect{r}_A ) = \phi_m ( \vect{d} ; \vect{\mu}_{d|r} , \matr{\Sigma}_{d|r} )
\end{align*}
with conditional expectation $ \vect{\mu}_{d|r} = \matr{H} \vect{r} $ where $ \matr{H} $ is an observation $ (m \times n) $-matrix.
Moreover, assume conditional independence of $ [ \vect{\nu} , \vect{ d} | \vect{r} ] $.

\noindent
Then the following holds:
\begin{description}
\item [A.] $ [ \vect{r}_A, \vect{d} ] $ is selection Gaussian
\item [B.] $ \vect{d} $ is selection Gaussian
\item [C.] $ [ \vect{r}_A | \vect{d} ] $ is selection Gaussian
\end{description}

\end{Prop}

\noindent
The Proposition is justified by the following Proof,

\begin{Proo}
The three proposition items are demonstrated sequentially.

\vspace{0.5cm}

\noindent
The joint pdf in \textbf{A} is,
\begin{align*}
[\vect{r}_A , \vect{d} ] &\rightarrow f ( \vect{r}_A , \vect{d} ) = f ( \vect{d} | \vect{r}_A ) f( \vect{r}_A ) \\
              &= \phi_m ( \vect{d} ; \vect{\mu}_{d|r}, \matr{\Sigma}_{d|r} ) \times                     
              \mbox{const}  \times \Phi_q ( \mathcal{A}_\nu ; \vect{\mu}_{\nu|r} , \matr{\Sigma}_{\nu|r} ) 
               \times \phi_n ( \vect{r} ; \vect{\mu}_r , \matr{\Sigma}_r )   \\                    
              &= \mbox{const} \times \Phi_q ( \mathcal{A}_\nu ; \vect{\mu}_{\nu|r} , \matr{\Sigma}_{\nu|r} )
       \times       \phi_{n+m} 
\left( 
\left[
\begin{array}{c}
\vect{r} \\  \vect{d}
\end{array} \right]
; \left[ \begin{array}{c} 
\vect{\mu}_r  \\  \matr{H} \vect{\mu}_r  
\end{array}  \right] , 
\left[
\begin{array}{cc}
 \matr{\Sigma}_r  &  \matr{\Sigma}_r \matr{H}^T \\
\matr{H}  \matr{\Sigma}_r & \matr{H} \matr{\Sigma}_r \matr{H}^T + \matr{\Sigma}_{d|r}  
\end{array}  \right]
 \right) \\
  &= \mbox{const} \times \Phi_q ( \mathcal{A}_\nu ; \vect{\mu}_{\nu|rd} , \matr{\Sigma}_{\nu|rd} )
     \times        \phi_{n+m} 
\left( 
\left[
\begin{array}{c}
\vect{r} \\  \vect{d}
\end{array} \right]
; \left[ \begin{array}{c} 
\vect{\mu}_r  \\  \matr{H} \vect{\mu}_r  
\end{array}  \right] , 
\left[
\begin{array}{cc}
 \matr{\Sigma}_r  &  \matr{\Sigma}_r \matr{H}^T \\
\matr{H}  \matr{\Sigma}_r & \matr{H} \matr{\Sigma}_r \matr{H}^T + \matr{\Sigma}_{d|r}  
\end{array}  \right]
 \right) 
\end{align*}
with the last identity from conditional independence of $ [ \vect{\nu} , \vect{d} | \vect{r} ] $.

Hence from Definition \ref{def:SGpdf} , the joint $ (n+m) $-vector $ [ \vect{r}, \vect{d} ] $ is selection Gaussian.

\vspace{0.5cm}

\noindent
The marginal pdf in \textbf{B} is,
\begin{align*}
\vect{d} &\rightarrow f ( \vect{d} ) = \int  f ( \vect{r}_A , \vect{d} ) d \vect{r} \\
     &= \mbox{const} \int \Phi_q ( \mathcal{A}_\nu ; \vect{\mu}_{\nu|rd} , \matr{\Sigma}_{\nu|rd} )
    \times       \phi_{n+m} 
\left( 
\left[
\begin{array}{c}
\vect{r} \\  \vect{d}
\end{array} \right]
; \left[ \begin{array}{c} 
\vect{\mu}_r  \\  \matr{H} \vect{\mu}_r  
\end{array}  \right] , 
\left[
\begin{array}{cc}
 \matr{\Sigma}_r  &  \matr{\Sigma}_r \matr{H}^T \\
\matr{H}  \matr{\Sigma}_r & \matr{H} \matr{\Sigma}_r \matr{H}^T + \matr{\Sigma}_{d|r}  
\end{array}  \right]
 \right)
d \vect{r} \\
&= \mbox{const}  \times \int \Phi_q ( \mathcal{A}_\nu ; \vect{\mu}_{\nu|rd} , \matr{\Sigma}_{\nu|rd} ) 
               \times \phi_n ( \vect{r} ; \vect{\mu}_{r|d} , \matr{\Sigma}_{r|d} ) d \vect{r}
\times \phi_m ( \vect{d} ; \vect{\mu}_d , \matr{\Sigma}_d ) \\
&= \mbox{const} \times \E_{r|d} \{ \Phi_q ( \mathcal{A}_\nu ; \vect{\mu}_{\nu|rd} , \matr{\Sigma}_{\nu|rd} ) \}
  \times \phi_m ( \vect{d} ; \vect{\mu}_d , \matr{\Sigma}_d ) \\
&= \mbox{const} \times \Phi_q ( \mathcal{A}_\nu ; \vect{\mu}_{\nu|d} , \matr{\Sigma}_{\nu|d} )
   \times \phi_m ( \vect{d} ; \vect{\mu}_d , \matr{\Sigma}_d )
\end{align*}
with the last identity from Result R2G.

Hence from Definition \ref{def:SGpdf}, the marginal $ m $-vector $ \vect{d} $ is selection Gaussian.

\vspace{0.5cm}

\noindent
The conditional pdf in \textbf{C} is,
\begin{align*}
[ \vect{r}_A | \vect{d} ] &\rightarrow f ( \vect{r}_A | \vect{d} ) 
= \mbox{const} \times \frac{ f ( \vect{r}_A , \vect{d} ) }{ f ( \vect{d} ) } \\
 &= \frac{  \mbox{const} \times \Phi_q ( \mathcal{A}_\nu ; \vect{\mu}_{\nu|rd} , \matr{\Sigma}_{\nu|rd} )
          \times     \phi_{n+m} 
\left( 
\left[
\begin{array}{c}
\vect{r} \\  \vect{d}
\end{array} \right]
; \left[ \begin{array}{c} 
\vect{\mu}_r  \\  \matr{H} \vect{\mu}_r  
\end{array}  \right] , 
\left[
\begin{array}{cc}
 \matr{\Sigma}_r  &  \matr{\Sigma}_r \matr{H}^T \\
\matr{H}  \matr{\Sigma}_r & \matr{H} \matr{\Sigma}_r \matr{H}^T + \matr{\Sigma}_{d|r}  
\end{array}  \right]
 \right) }{ \mbox{const} \times \Phi_q ( \mathcal{A}_\nu ; \vect{\mu}_{\nu|d} , \matr{\Sigma}_{\nu|d} )
   \times \phi_m ( \vect{d} ; \vect{\mu}_d , \matr{\Sigma}_d ) } \\
&= \mbox{const} \times \Phi_q ( \mathcal{A}_\nu ; \vect{\mu}_{\nu|rd} , \matr{\Sigma}_{\nu|rd} ) 
\times \phi_n ( \vect{r} ; \vect{\mu}_{r|d} , \matr{\Sigma}_{r|d} )  
\end{align*}
Hence from Definition \ref{def:SGpdf}, the conditional $ n $-vector $ [ \vect{r}_A | \vect{d} ] $ is selection Gaussian.

\end{Proo}

%
%
%
%

\newpage
\section*{Supplementary Material}
\setcounter{section}{0}

%
%
%
%

\section{Sampling - truncated multivariate Gaussian distribution} \label{sec:app_sample}

Consider the problem of sampling from a $n$-dimensional truncated multivariate normal distribution with unnormalized density $I(\mathbf x \in A) \times \phi_n(\mathbf x ; \boldsymbol \mu, \boldsymbol \Sigma)$, where $\mathbf x, \boldsymbol \mu \in \mathcal R^n$, $ \boldsymbol \Sigma \in \mathcal R^{n \times n}$, $A = A_1 \times \ldots \times A_n$, $A_i \subseteq \mathcal R$, $I(\cdot)$ is the indicator function, and $\phi_n(\mathbf x ; \boldsymbol \mu, \boldsymbol \Sigma)$ is the multivariate Gaussian density distribution with expectation vector $\boldsymbol \mu$ and covariance matrix $\boldsymbol \Sigma$. In order to sample from this distribution we extend the Metropolis-Hastings algorithm in \citet{Robert:TruncGaussian} with a block independent proposal distribution:
\begin{align*}
p^*(\mathbf x^a \mid \mathbf x^b)
&= \prod_{i=1}^q I( x_i^a \in A_i) \; \frac{\phi_1(x_i^a \mid \mathbf x_{1:i-1}^a, \mathbf x^b ; \boldsymbol \mu, \boldsymbol \Sigma)}{\Phi_1( x^a_i \in A_i \mid \mathbf x_{1:i-1}^a,\mathbf x^b ; \boldsymbol \mu, \boldsymbol \Sigma)}
, \label{eqn:mcmcproposal}
\end{align*}
where $n_a$ is the block size, $\mathbf x^a \in \mathcal{R}^{n_a}, \mathbf x^b \in \mathbb{R}^{n-n_a}$, $\phi_1(x_i^a \mid \mathbf x_{1:i-1}^a, \mathbf x^b ; \boldsymbol \mu, \boldsymbol \Sigma)$ the conditional Gaussian probability of $x_i^a$ given $\mathbf x_{1:i-1}^a$ and $\mathbf x^b$, and $\Phi_1( x^a_i \in A_i \mid \mathbf x_{1:i-1}^a, \mathbf x^b  ; \boldsymbol \mu, \boldsymbol \Sigma)$ is the probability of the set $A_i$ under the Gaussian probability distribution of $x_i$ given $\mathbf x_{1:i-1}^a$ and $\mathbf x^b$. We use the notation $\mathbf x_{1:i-1} = (x_1,x_2, \ldots x_{i-1})$. The approach is inspired by the importance sampler in \citet{Genz1992}. Note that $p^*(\mathbf x^a \mid \mathbf x^b)$ is normalized and it is easy to sample from the distribution due to the sequential structure. 

The acceptance probability in the accept/reject step is
\begin{align*}
\alpha 
& = \min \left\lbrace 1, \frac{p({\mathbf x^a}' \mid \mathbf x^b)}{p(\mathbf x^a \mid \mathbf x^b)} \cdot \frac{p^*(\mathbf x^a \mid \mathbf x^b)}{p^*({\mathbf x^a}' \mid \mathbf x^b)} \right\rbrace \notag \\ 
& = \min \left\lbrace 1, \frac{ \prod_{i=1}^{n_a} \Phi_1({ x^a_i}' \in A_i \mid {\mathbf x^a_{1:i-1}}', \mathbf x^b  ; \boldsymbol \mu, \boldsymbol \Sigma)}{ \prod_{i=1}^{n_a} \Phi_1(x^a_i \in A_i \mid \mathbf x^a_{1:i-1}, \mathbf x^b  ; \boldsymbol \mu, \boldsymbol \Sigma)} \right\rbrace,
\end{align*}
where ${\mathbf x^a}'$ is the new proposed state.
The Metropolis-Hastings algorithm is presented in Algorithm \ref{alg1}.

\begin{algorithm}[H]
\DontPrintSemicolon
\BlankLine\;
Initialize $\mathbf x$ with a value in $A$. \;
\textbf{Iterate} \;
\quad Choose one element $i$ at random in $\mathbf x$.  \;
\quad Find the set of the $n_a$ closest by correlation element to $i$. \; 
\quad Define the set of the $n_a$ elements $a_i$ and $b_i$ as it complement. \;
\quad Sample $\mathbf x'_{a_i \mid b_i} \sim p^*(\mathbf x^{a_i} \mid \mathbf x^{a_i})$. \;
\quad Accept $\mathbf x'_{a_i \mid b_i}$ with probability $\alpha$. \;
\textbf{End} \;
\BlankLine \;
\caption{Sampling from truncated multivariate normal distribution}
\label{alg1}
\end{algorithm}

In practice we calculate the conditional distributions in Algorithm \ref{alg1} in advance. To save memory and time we also limit the elements in $\mathbf x$, i.e. sets eligible for choice, such that all elements in $\mathbf x$ has approximately equal update probability. We normally use the block size $n_a = 100$.

\section{Estimation - multivariate Gaussian probabilities} \label{sec:app_is}

Consider the problem of estimating the multivariate Gaussian probability 
\begin{align*}
\Phi_n( A;\boldsymbol \mu, \boldsymbol \Sigma)
&= \int I(\mathbf x \in A) \; \phi_n(\mathbf x;\boldsymbol \mu, \boldsymbol \Sigma) \; \mathrm d \mathbf x,
\end{align*} 
where $\mathbf x, \boldsymbol \mu \in \mathcal R^n$, $ \boldsymbol \Sigma \in \mathcal R^{n \times n}$, $A = A_1 \times \ldots \times A_n$, $A_i \subset \mathcal R$, $I(\cdot)$ is the indicator function,  and $\phi_n(\mathbf x ; \boldsymbol \mu, \boldsymbol \Sigma)$ is the multivariate Gaussianl density distribution with expectation vector $\boldsymbol \mu$ and covariance matrix $\boldsymbol \Sigma$.
The usual importance sampling Monte Carlo approximation is 
\begin{align*}
\Phi_n(A;\boldsymbol \mu, \boldsymbol \Sigma) 
&\approx \sum_{j=1}^N I(\mathbf x^j  \in A) \; \frac{\phi_n(\mathbf x^j;\boldsymbol \mu, \boldsymbol \Sigma)}{f_n(\mathbf x^j;\boldsymbol \mu, \boldsymbol \Sigma)},
\end{align*}
with $\mathbf x^j \sim f_n(\mathbf x;\boldsymbol \mu, \boldsymbol \Sigma); \; j = 1,\ldots N$ and $N$ is the number of Monte Carlo sampling points. We extend the approach presented in \citet{Genz1992} by allowing $A_i$ to consist of several intervals, and use
\begin{align*}
f_n(\mathbf x;\boldsymbol \mu, \boldsymbol \Sigma)
&= \prod_{i=1}^n I(x_i \in A_i) \; \frac{\phi_1(x_i \mid \mathbf x_{1:i-1} ; \boldsymbol \mu, \boldsymbol \Sigma)}{\Phi_1(A_i \mid \mathbf x_{1:i-1} ; \boldsymbol \mu, \boldsymbol \Sigma)},
\end{align*}
as importance function, where $\phi_1(x_i\mid \mathbf x_{1:i-1}; \boldsymbol \mu, \boldsymbol \Sigma)$ the conditional Gaussian probability of $x_i$ given $\mathbf x_{1:i-1}$, and $\Phi_1( A_i \mid \mathbf x_{1:i-1}; \boldsymbol \mu, \boldsymbol \Sigma)$ is the probability of the set $A_i$ under the Gaussian probability distribution of $x_i$ given $\mathbf x_{1:i-1}$. We use the notation $\mathbf x_{1:i-1} = (x_1,x_2, \ldots x_{i-1})$. However, we also introduce a mean shift parameter $\boldsymbol \eta$ in the importance function which is important for asymmetric sets $A_i$. Then the importance sampling approximation appear as
\begin{align*}
\Phi_q(A;\boldsymbol \mu, \boldsymbol \Sigma) 
& \approx \sum_{j=1}^N \frac{\phi_n(\mathbf x^j;\boldsymbol \mu, \boldsymbol \Sigma)}{\phi_n(\mathbf x^j;\boldsymbol \mu + \boldsymbol \eta, \boldsymbol \Sigma)} \prod_{i=1}^n \Phi_1(A_i \mid \mathbf x^j_{1:i-1} ; \boldsymbol \mu + \boldsymbol \eta, \boldsymbol \Sigma), 
\end{align*}
with $\mathbf x^j \sim f_n(\mathbf x;\boldsymbol \mu+ \boldsymbol \eta, \boldsymbol \Sigma), \; j = 1,\ldots N$.

\section{Example: Prior Model} \label{sec:app_exPrior}

The flexibility of the selection Gaussian pdf as prior model for the spatial variable of interest is demonstrated by generating realizations with varying model parameter sets, $ \vect{\theta}_p = ( \mu, \sigma^2, \gamma, \rho(\vect{ \tau}), \mathcal{A}_i) $. The spatial variable is represented on a $ (64 \times 64) $-grid $ \mathcal{L}_{\mathcal{D}} $ covering $ \mathcal{D} \subset \mathcal{R}^2 $, hence a surface in two dimensions. The spatial correlation function is parametrized as $ \rho ( \vect{\tau}; (d_h , d_v )) = \exp \{ - [[ \tau_h^2 / d_h^2 ]
+ [ \tau_v^2 / d_v^2 ]] \} $, hence to be a second-order exponential correlation function with anisotropy factors $ (d_h , d_v ) $. The selection set $ \mathcal{A}_i \subset \mathcal{R} $ is parametrized as a number of line segments on $ \mathcal{R} $. The example design is summarized in Table \ref{tbl:app_synthetic_param}.

\vspace{0.5cm}

\begin{table}
\caption{Model parameters for six cases, with $\mu = 0 $ and  $\sigma^2=1$ for all cases.}
\begin{center}
\begin{small}
\begin{tabular}{c|ccccl }
Case & $\gamma$ & $d_h$ & $d_v$ & $A_i$ & description \\ \hline
1    & 0.8000   & 2.0   & 2.0  & $(\infty, -0.3] \cup [0.3, \infty)$ & sym. bimodal iso. \\ 
2    & 0.6500   & 6.0   & 0.85  & $(\infty, -0.3] \cup [0.3, \infty)$ & asym. bimodal aniso. \\
3    & 0.9250   & 2.0   & 0.60  & $(\infty, -0.85] \cup [0.8, \infty)$& sym. bimodal aniso. \\
4    & 0.9995   & 3.0   & 3.0  & $[-0.45, -0.2] \cup [-0.1, 0.1] \cup [0.2, 0.45]$ & sym. trimodal iso.\\
5    & 0.7000   & 2.0   & 2.0  & $(\infty, -0.7] \cup [-0.1, 2.5]$    & asym. unimodal iso. \\
6    & 0.7000   & 2.0   & 2.0  & $(\infty, -1.75] \cup [-0.5, 0.5] \cup [1.75, \infty)$ & sym. heavy tailed iso. \\
\end{tabular}
\end{small}
\end{center}
\label{tbl:app_synthetic_param}
\end{table}

In order to simulate realizations from the selection Gaussian pdf we extend the Metropolis Hastings (MH) algorithm presented in \citet{Rimstad2012} by allowing more general selection sets $ \mathcal{A}$. The algorithm is summarized in Supplement \ref{sec:app_sample}. The algorithm is a block proposal MH-algorithm, and we normally use block sizes about $100$ which in our examples give an acceptance rate of about $0.25$. The computer demand for generating one realization is a couple of minutes on a regular laptop computer. The burn-in and mixing appear as satisfactory and are not displayed.

\begin{figure}
 \centering
\includegraphics[height=0.7\textheight]{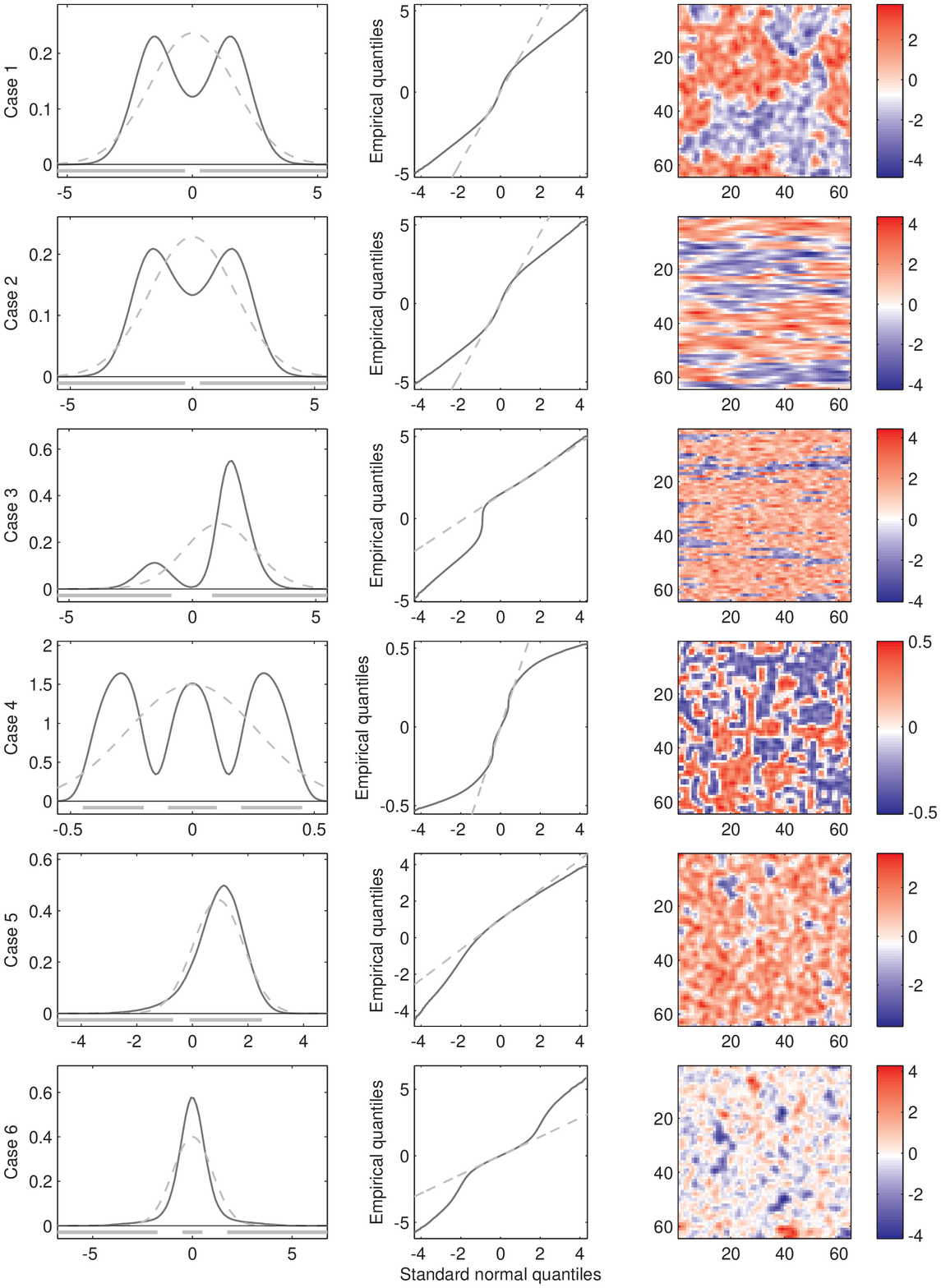}
 \caption{First column: marginal distribution of selection Gaussian random field ( solid black), standard normal distribution ( dashed gray), and selection sets on auxiliary random field on axis ( solid gray). Second column: quantile-quantile plot of marginal selection Gaussian random field versus theoretical quantiles from the Gaussian distribution. Third column: realization from selection Gaussian random field.}
 \label{fig:app_synthetic_marginals}
\end{figure}

Figure \ref{fig:app_synthetic_marginals} displays the results from the six cases. The first column displays the marginal distribution in the $(32, 32)$ location of the grid compared with a univariate Gaussian distribution with the same two first moments as the marginal selection Gaussian pdf. The selection set $ \mathcal{A} $  is illustrated with a thick gray line at the bottom of the display. The selection set is comparable to the marginal of the selection Gaussian pdf  because  $\sigma^2=1$. The second column displays Gaussian quantile-quantile plots of the marginal distributions. The last column displays realizations from the selection Gaussian prior pdfs.

The first row in Figure \ref{fig:app_synthetic_marginals}, case 1, displays a symmetric bimodal spatially isotropic model. The selection region for the auxiliary variable is absolute values greater than $0.3$. The marginal pdf is symmetric and bimodal, and the quantile-quantile plot shows clear deviations from the Gaussian distribution. In the realization the two modes are visible as two separated levels with sharp transitions between them.

Case 2 is displayed in the second row in Figure \ref{fig:app_synthetic_marginals} and this model is also symmetric and bimodal, but  spatially anisotropic. The  selection region for the auxiliary variable is absolute values greater than $0.3$, as in case 1. In this case the horizontal spatial correlation is increased and the vertical one decreased, while the coupling parameter is reduced. The resulting realization is clearly layered, with marginal pdf very similar to the one in  case 1.

The third row in Figure \ref{fig:app_synthetic_marginals}, case 3, displays an asymmetric bimodal spatially anisotropic model. The selection is asymmetric and further out in the tails than for previous cases. The occurrence of two clearly separated modes is made  possible by  low spatial correlation which allows larger jumps. The asymmetric selection causes the mode to the left to be smaller than the mode to the right. In spite of the low correlation the realization appears with clear spatial anisotropy  and with two distinct modes. 

Case 4 is displayed in the forth row in Figure \ref{fig:app_synthetic_marginals} and represents a symmetric trimodal spatial isotropic model. The selection contains  three symmetric closed intervals, which provides a trimodal symmetric marginal pdf. The three modes are distinctly separated, and clearly visible in the realizations. The spatial transitions  between  the two outer modes seem always to pass through the middle mode.

The fifth row in Figure \ref{fig:app_synthetic_marginals}, case 5, displays a skewed model. The skewed models considered in \citet{Allard2007} and \citet{Rimstad2012} only contain one-sided selection interval, hence the model formulation constrains the degree of skewness. In the current case one additional selection interval is introduced, which provides a more flexible skewness structure for the model. The skewness is evident in the marginal pdf, in the quantile-quantile plot, and in the realization of the selection Gaussian prior pdf.

The last row in Figure \ref{fig:app_synthetic_marginals}, case 6, displays a symmetric peaked heavy tailed model. Symmetric selection is used, which forces higher probability density in the centre and in the tails. The extreme tails still decay exponentially, as  seen in the quantile-quantile plot, while the more visible effects is due to the heavy tails. The closest univariate Student-$t$ distribution, if we ignore the extreme tails, has about $2$ degrees of freedom.

\section{Example: Posterior Model} \label{sec:app_exPosterior}

The posterior model and the effect of using different prediction criteria is demonstrated on a small example.
The spatial variable is defined on a 128-grid $ \mathcal{L}_\mathcal{D} $ covering $ \mathcal{D} \subset \mathcal{R} $, hence a one-dimensional case. The prior from the selection Gaussian class is defined with the spatial correlation function $ \rho( \tau) = \exp \{ - \tau^2 \} $, and a selection set $ \mathcal{A} \subset \mathcal{R} $ consisting of several line segments. The example design is summarized in  Table \ref{tbl:app_synthetic_param_prediction}.  The four cases shear about the same characteristics as case $1, 4, 5$ and $6$ in the example in Supplement \ref{sec:app_exPrior}. The likelihood function provides exact observations at grid locations 16 and 112. The actual observed values are also specified in  Table \ref{tbl:app_synthetic_param_prediction}.

\begin{table}
\caption{Model parameters for four posterior cases, with $\mu = 0 $ and  $\sigma^2=1$ for all cases.}
\begin{center}
\begin{small}
\begin{tabular}{c|ccccc }
Case & $\gamma$ & $d_h$ & $A_i$ & description & cond. values\\ \hline
1    & 0.900  & 4   & $(\infty, -0.4] \cup [0.4, \infty)$ & sym. bimodal & $ 2.5, -2.5$\\ 
2    & 0.999  & 4   & $[-0.65, -0.4] \cup [0.12, 0.12] \cup [0.4 0.65]$    & sym. trimodal & $ 0.55, -0.55$\\
3    & 0.600  & 4   & $(\infty, -1.5] \cup [-0.5, 0.5)$ & asym. unimodal & $1.0, -3.0$\\
4    & 0.700 & 4   & $(\infty, -1.75] \cup [-0.5, 0.5] \cup [1.75, \infty)$& sym. heavy tailed & $ 3.0, -3.0$\\
\end{tabular}
\end{small}
\end{center}
\label{tbl:app_synthetic_param_prediction}
\end{table}

The posterior model is from the selection Gaussian class, and the exact observations are of course exactly reproduced. The
corresponding conditional selection Gaussian pdf must be calculated and realizations from this pdf can be generated by
the algorithm in Supplement \ref{sec:app_sample}.

\begin{figure}
 \centering
\includegraphics[width=0.7\textwidth]{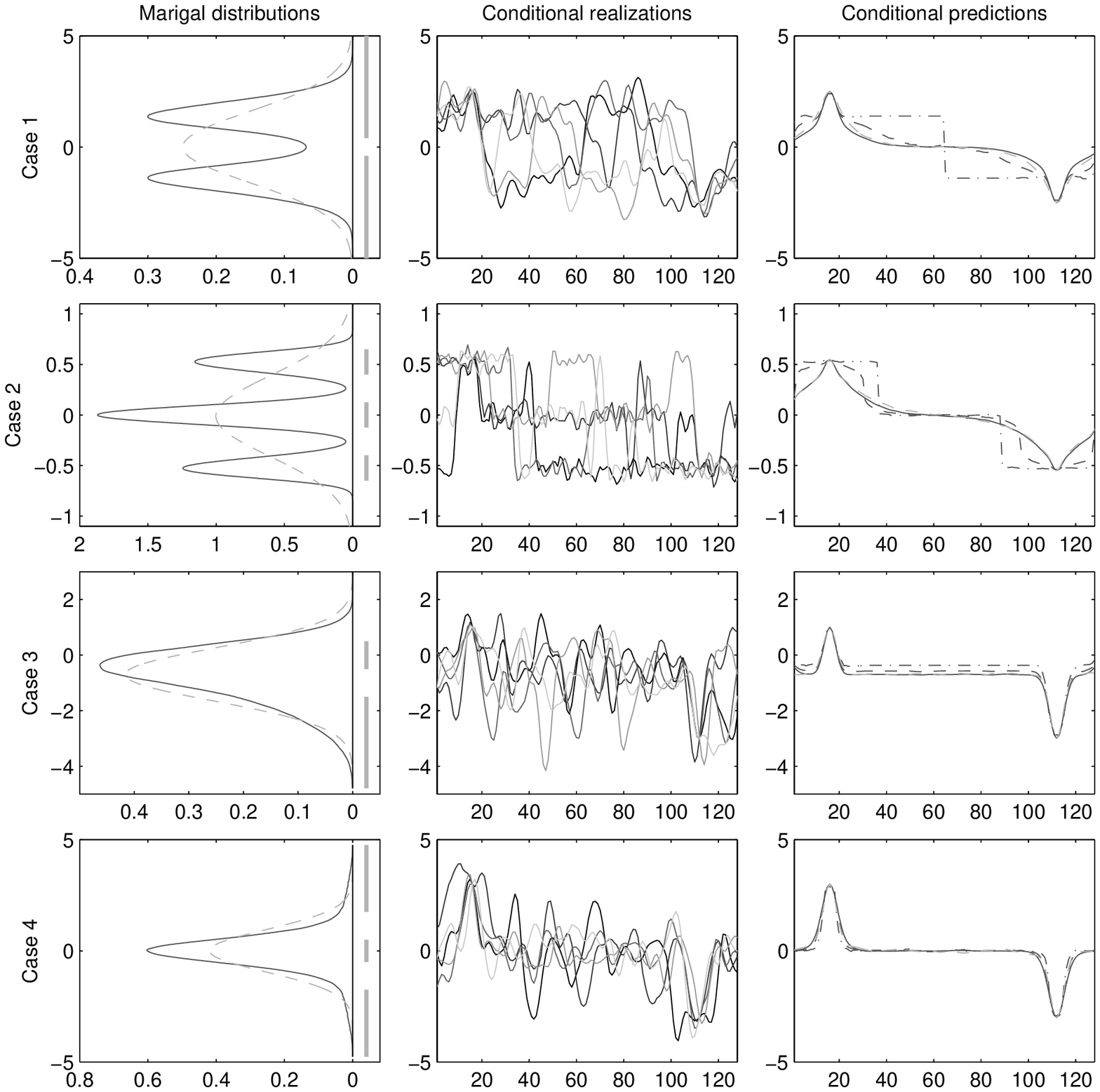} 
\caption{First column: marginal distribution of prior selection Gaussian model ( solid black) and corresponding Gaussian model ( dashed gray), and selection set on auxiliary random field on axis (solid gray). Second column: five realizations of the posterior selection Gaussian random field. Third column: posterior selection Gaussian model predictions, with  E-prediction ( solid black), MED-prediction ( dashed black), and MAP-prediction (dashed-dotted black). The corresponding Gaussian model prediction (E/MED/MAP)  ( dashed gray).}
 \label{fig:app_synthetic_prediction}
\end{figure}

We compare Bayesian inversion based on selection Gaussian and Gaussian prior models, and the effect of using different predictors. The two first moments are set identical in the two alternative  prior models for each case. The results are summarized in Figure \ref{fig:app_synthetic_prediction}. The first column in the figure contains the marginal pdfs at location $ 64 $ for both the selection Gaussian and Gaussian prior models for the four cases. Selection sets for the former are plotted as grey line segments. The second column contain realizations from the posterior pdf of the selection Gaussian model given observations at locations $16$ and $112$.  Lastly, the third column contains predictions based on different criteria for each of the two posterior models for each case.  The E (black solid), MED (black hatched) and MAP (black hatch-dot) predictors for the selection Gaussian model and the E/MED/MAP (grey hatched) predictors,
which coinsides, for the Gaussian model. We focus our discussion on the results from the selection Gaussian model, and compare them with the Gaussian predictor.

The first row, case 1, displays a symmetric bi-modal prior model. The marginal distribution in Figure \ref{fig:app_synthetic_prediction} is clearly bimodal. We condition on the values $2.5$ and $-2.5$ at grid nodes $16$ and $112$, respectively. The realizations of the posterior model have a evident bi-modal structure. The E predictor is almost identical to the Gaussian predictor, while the MED and MAP predictors clearly deviate from the Gaussian predictor. The MAP predictor has a stepwise structure reflecting the bimodality of the posterior model and the MED predictor is somewhere between the MAP and E predictors, but closest to the E predictor.

Case 2, displayed in the second row,  has a symmetric tri-modal prior model. We condition on the values $0.55$ and $-0.55$ at grid nodes $16$ and $112$, respectively. The three modes are clearly visible in both the marginal distribution of the prior and the posterior realizations. The E predictor is in this case also almost identical to the Gaussian predictor. The MAP predictor has a stepwise structure with three levels, and the MED predictor is in this case closest to the MAP predictor.

The third row, case 3, displays an asymmetric unimodal prior model. We condition on the values $1.0$ and $-3.0$ at grid nodes $16$ and $112$, respectively. The marginal distribution of the prior is obviously skewed and the posterior realizations have a skewed structure. The E predictor and the Gaussian predictor are again almost identical. The MAP and MED predictors are similar to the E predictor except that the stationary levels for the MAP and MED are somewhat shifted relative to the E predictor.

Case 4,  displayed in the last row, has a symmetric heavy tailed prior model. We condition on the values $3.0$ and $-3.0$ at grid nodes $16$ and $112$, respectively. All the predictions have similar shapes, but the MAP predictor, followed by the MED predictor, decays faster toward the stationary level than the E predictor. Again the E predictor and the Gaussian predictor are almost identical. The fact that the E, MED and MAP predictors are not identical entails that the posterior distributions are asymmetric, in spite the prior distribution being unimodal and symmetric.

The E, MED and MAP predictors can be very different for selection Gaussian models, contrary to the Gaussian model where all the three predictors are identical. The predictors are particularly different for  multi-modal prior models, where the MAP predictor appears as stepwise. The E predictors for the selection Gaussian model is almost identical to the predictor for the corresponding Gaussian prior model.

\section{Example: Model Parameter Inference} \label{sec:app_exInference}

 We consider case 1 in the example in Supplement \ref{sec:app_exPrior}, which has a selection Gaussian prior model with symmetric, bi-modal marginal pdfs and isotropic spatial correlation. The prior model is parametrized by $ \vect{\theta}_p = ( \mu, \sigma^2, d, \gamma, a) $ , where $ ( \mu, \sigma^2, d) $ defines the stationary, isotropic Gaussian basis-pdf with the two former parameters being expectation and variance respectively, while the latter is the range in an isotropic second-order exponential spatial correlation function. The parameter $ \gamma $ is the coupling parameter and $ a $ defines the selection set $ \mathcal{A}_i : (- \infty, -a] \cup [a, \infty) \subset \mathcal{R} $.
 
 Consider one training image $ \vect{r}_A^o $ discretized to grid $ \mathcal{L}_\mathcal{D} $. We estimate the prior model parameters $ \vect{\theta}_p $ by a maximum likelihood approach. The challenging calculations are repeated assessment of  $\log \Phi_p(\mathcal{A}; \mathbf 0, (1-\gamma^2) \mathbf I_p + \gamma^2 \mathbf C)$ for varying values of $ \vect{\theta}_p $, which we solve by Monte Carlo importance sampling approach inspired by  \citet{Genz1992} and \citet{Genz2009}, see Supplement \ref{sec:app_is}. For a grid $ \mathcal{L}_\mathcal{D} : [32 \times 32] $ the computation of the likelihood function for one set of $ \vect{\theta}_p $ requires about one minute on a regular laptop computer. In order to ensure a smooth likelihood function which can be optimized by standard procedures we keep the Monte Carlo samples fixed during each optimization. We used $ N= 5000 $ samples which according to results in  \citet{Rimstad2012} should ensure stable solutions. There may exist multiple local optima in the likelihood function, hence we initiate the optimization in multiple points, but we encountered few problems with multiple solutions.
 
 We used the following experimental design: generate $ 1000 $ realizations of the model with parameters  $ \vect{\theta}_p = ( 0, 1, 2, 0.8, 0.3) $ on a grid $ \mathcal{L}_\mathcal{D} : [32 \times 32] $. For each realization select one subset on grid sizes: $ [8 \times 8] , [16 \times 16], [24 \times 24] $ and $ [32 \times 32] $, and estimate the model parameter $ \vect{\theta}_p $ as previously described.
 
\begin{figure}
 \centering
\includegraphics[width=0.55\textwidth]{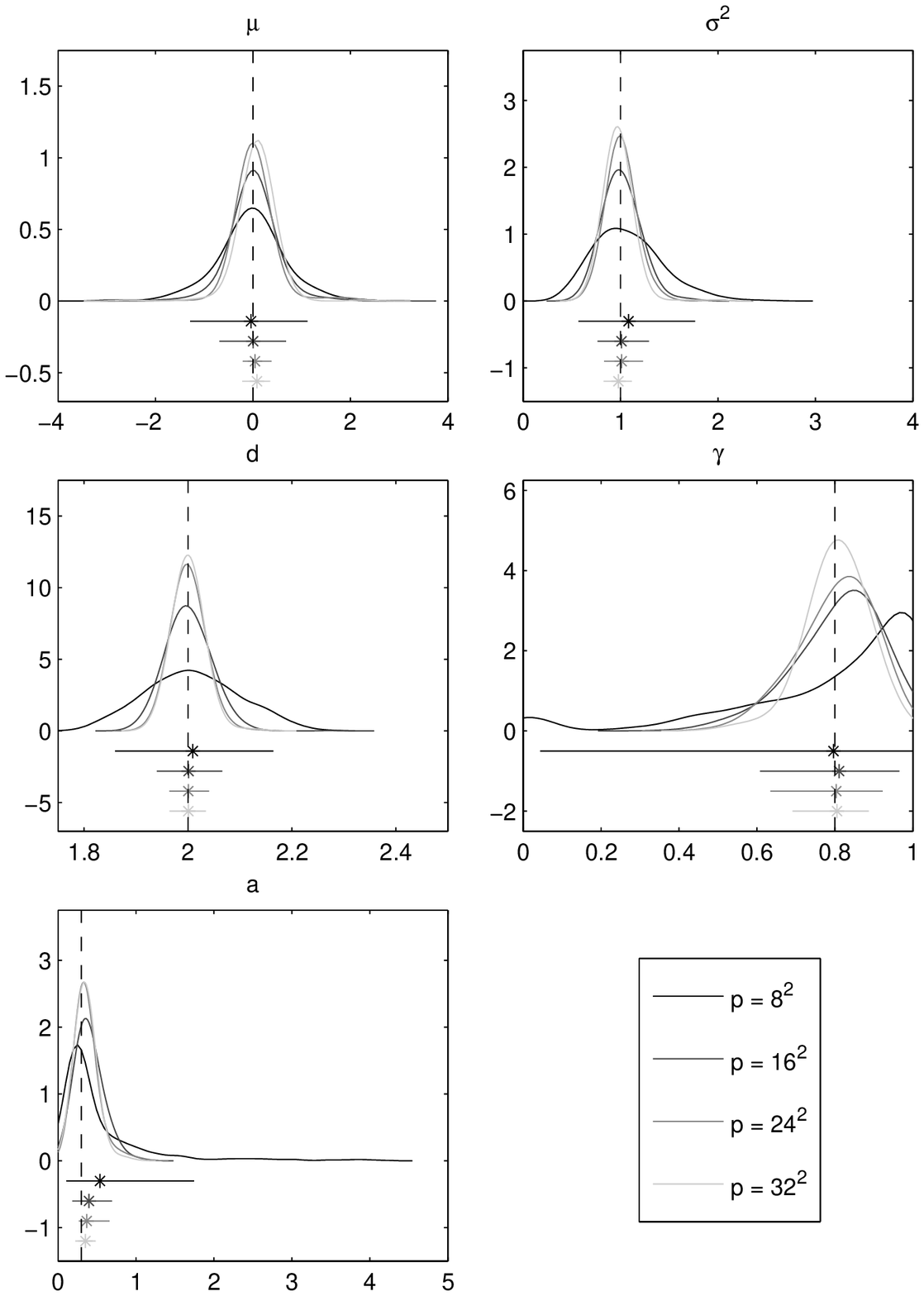}
 \caption{Density plots of parameter estimates $\hat{ \vect{\theta}}_p $ with increasing size of the training image $ \vect{r}^o $. Below are means and 90\% confidence intervals, and true values ( vertical dashed lines).}
 \label{fig:app_para_est}
\end{figure}

\begin{figure}
 \centering
\includegraphics[width=1.0\textwidth]{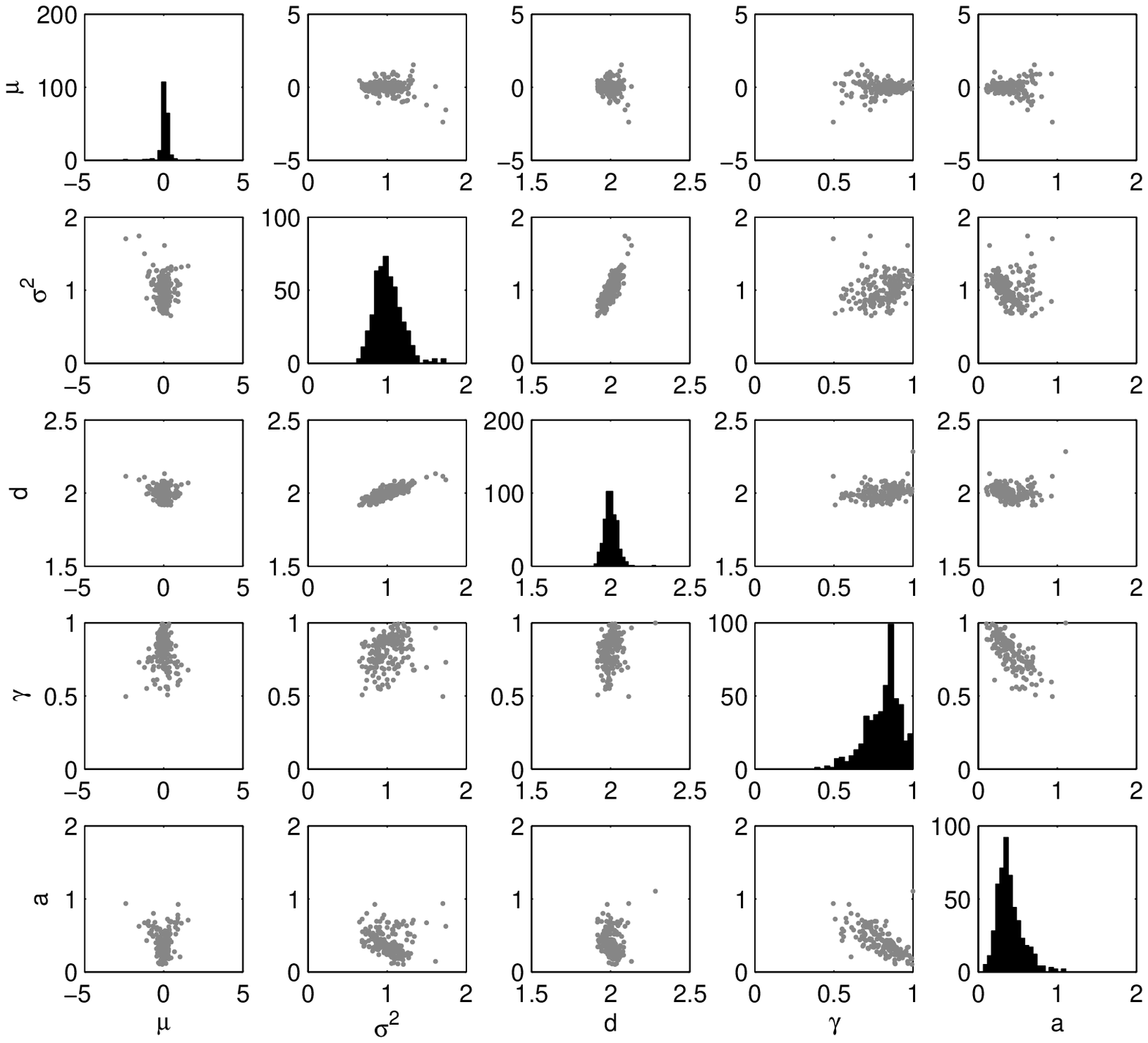}
 \caption{Cross-plot of the estimated parameters $\hat{ \vect{\theta}}_p $ for size $p = 16^2$. }
 \label{fig:app_para_est_cross}
\end{figure}

 The results are summarized in Figure \ref{fig:app_para_est} and \ref{fig:app_para_est_cross}. From the Figure \ref{fig:app_para_est} we observe that the estimator for $ \vect{\theta}_p $ is relatively well centred at the correct values even for small grid sizes. The centering improves with increasing grid sizes, while the estimation variance
decreases. Hence it appears as the estimator is consistent with increasing grid sizes, but not in general unbiased. These results are as expected for maximum likelihood estimators. In Figure \ref{fig:app_para_est_cross} we display multivariate results for the grid size $ [16 \times 16] $. The negative correlation between the estimators for $ \gamma $ and $ a $, the coupling and selection set, is easy to understand since stronger coupling requires selections closer to zero. Moreover, there appears to be some positive correlation between estimators for $ \sigma^2 $ and $ d $, variance and range.

It is encouraging that relatively stable parameter estimates can be obtained from small training images of 
grid size about $ [24 \times 24] $.
These results make us believe that the parametrization of the selection Gaussian prior model is reasonable and that the model parameters can be robustly assessed from training images of reasonable size. Moreover, it makes us trust the numerical approximations used in the optimization procedure. 
 Cases 2 through 6 in the example in Supplement \ref{sec:app_exPrior}, require parametrizations with more parameters, which may complicate the evaluations of the likelihood functions and introduce ambiguities among parameters. These complications are not further considered in the current study. 

We have based our inference study on a training image of exact, complete set of observations on a grid. Alternatively, maximum likelihood inference can be made from any set of observations from a Gauss-linear likelihood model since the posterior model will also be a selection Gaussian pdf. 

\end{document}